\documentclass[journal]{IEEEtran}
\usepackage{amsmath,graphicx}
\usepackage{cite}
\usepackage{array}
\usepackage{amssymb}
\usepackage{subfigure}
\usepackage{flushend}
\usepackage{stfloats}
\usepackage{dsfont}
\usepackage{bm}
\usepackage[caption=false,font=footnotesize]{subfig}
\usepackage{url}
\usepackage{empheq}
\usepackage[nodisplayskipstretch]{setspace}

\newtheorem{lemma}{{Lemma}}
\newtheorem{assumption}{{ Assumption}}
\newtheorem{theorem}{{Theorem}}
\newtheorem{corollary}{{Corollary}}

\newcommand{\bp}{ \begin{proof}}
\newcommand{\ep}{\end{proof} }

\newcommand{\Ex}{\mathbb{E}\hspace{0.05cm}}
\newcommand{\be}{\begin{equation}}
\newcommand{\ee}{\end{equation}}
\newcommand{\ben}{\begin{equation*}}
\newcommand{\een}{\end{equation*}}
\newcommand{\bal}{\begin{align}}
\newcommand{\eal}{\end{align}}
\newcommand{\bq}{\begin{eqnarray}}
\newcommand{\eq}{\end{eqnarray}}
\newcommand{\bqn}{\begin{eqnarray*}}
\newcommand{\eqn}{\end{eqnarray*}}
\newcommand{\nn}{\nonumber \\}
\newcommand{\diag}{\mbox{\rm diag}}
\def\w{{\boldsymbol{w}}}

\def\Zint{{\mathchoice{\setbox1=\hbox{\sf Z}\copy1\kern-.75\wd1\box1}
{\setbox1=\hbox{\sf Z}\copy1\kern-.75\wd1\box1}
{\setbox1=\hbox{\scriptsize\sf Z}\copy1\kern-.75\wd1\box1}
{\setbox1=\hbox{\scriptsize\sf Z}\copy1\kern-.75\wd1\box1}}}

\makeatletter
\def\hlinewd#1{%
  \noalign{\ifnum0=`}\fi\hrule \@height #1 \futurelet
   \reserved@a\@xhline}
\makeatother

\begin{document}
%
\title{Coordinate-Descent Diffusion Learning by Networked Agents}
\author{Chengcheng~Wang,~\IEEEmembership{Student~Member,~IEEE}, Yonggang~Zhang,~\IEEEmembership{Senior~Member,~IEEE}, Bicheng~Ying,~\IEEEmembership{Student~Member,~IEEE}, and Ali~H.~Sayed,~\IEEEmembership{Fellow,~IEEE}\thanks{A short version of this work appears in the conference publication \cite{wang2017}.}
\thanks{This work was performed while C. Wang was a visiting student at the UCLA Adaptive Systems Laboratory. The work of C. Wang was supported in part by a Chinese Government Scholarship. The work of Y. Zhang was supported in part by the National Natural Science Foundation of China (61371173), and Fundamental Research Founds for the Central University of Harbin Engineering University (HEUCFP201705). The work of B. Ying and A. H. Sayed was supported in part by NSF grants CCF-1524250 and
ECCS-1407712. }
\thanks{C. Wang and Y. Zhang are with the College of Automation, Harbin Engineering University, Harbin, Heilongjiang 150001 China. C. Wang is also with the School of Electrical and Electronic Engineering, Nanyang Technological University, 639798 Singapore (e-mail: \protect\url{wangcc@ntu.edu.sg}; \protect\url{zhangyg@hrbeu.edu.cn}).}
\thanks{B. Ying is with the Department of Electrical Engineering, University
of California, Los Angeles, CA 90024 USA (e-mail: \protect\url{ybc@ucla.edu}).}
\thanks{A. H. Sayed is with the Ecole Polytechnique Federale de Lausanne,
EPFL, School of Engineering, CH-1015 Lausanne, Switzerland (e-mail: \protect\url{ali.sayed@epfl.ch}).}}
\maketitle

\begin{abstract}
This work examines the mean-square error performance of diffusion stochastic algorithms under a generalized coordinate-descent scheme. In this setting, the adaptation step by each agent is
limited to a random subset of the coordinates of its stochastic gradient vector. The selection of
coordinates varies randomly from iteration to iteration and from agent to agent across the network. Such
schemes are useful in reducing computational complexity at each iteration in power-intensive large data applications. They are also useful in modeling situations where some partial gradient information may be missing at random. Interestingly, the results show that the steady-state performance of the learning strategy is not always degraded, while the convergence rate suffers some degradation. The results provide yet another indication of the resilience and robustness of adaptive distributed strategies.
\end{abstract}
\section{Introduction and Related Work}
Consider a strongly-connected network of $N$ agents, where information can flow in either direction between any two connected agents and, moreover, there is at least one self-loop in the topology \cite[p. 436]{ASayed2014}. We associate a strongly convex differentiable risk, $J_k(w)$, with each agent $k$ and assume in this work that all these costs share a common minimizer, $w^o\in\mathbb{R}^{M}$, where $\mathbb{R}$ denotes field of real numbers. This case models important situations where agents work cooperatively towards the same goal. The objective of the network is to determine the unique minimizer $w^o$ of the following aggregate cost, assumed to be strongly-convex:
\begin{equation}\label{equ1}
J^{\rm glob}(w)\triangleq\sum\limits_{k=1}^{N}J_k(w)
\end{equation}
It is also assumed that the individual cost functions, $J_k(w)$, are each twice-differentiable and satisfy
\begin{equation}\label{equ6}
0<{\nu_d}I_{M}\leq\nabla_w^2J_k(w)\leq{\delta_d}I_{M}
\end{equation}
where $\nabla^2_wJ_k(w)$ denotes the $M\times M$ Hessian matrix of $J_k(w)$ with respect to $w$, $\nu_d\leq\delta_d$ are positive parameters,  and $I_{M}$ is the $M\times M$ identity matrix. In addition, for matrices $A$ and $B$, the notation $ A\leq B$ denotes that $B-A$ is positive
semi-definite. The condition in (\ref{equ6}) is automatically satisfied by important cases of interest, such as logistic regression or mean-square-error designs \cite{ASayed2014,Sayed2014}.

Starting from some initial conditions $\{\bm w_{k,-1}\}$, the agents work cooperatively in an adaptive manner to seek the minimizer $w^o$ of problem (\ref{equ1}) by applying the following diffusion strategy \cite{ASayed2014,Sayed2014}:
\begin{subequations}\label{equ2}
\begin{empheq}[left=\empheqlbrace]{align}
\bm \phi_{k,i-1} &= \displaystyle{\sum_{\ell\in\mathcal{N}_k}}a_{1,\ell k}\bm w_{\ell,i-1}\label{equ2c}\\
\bm \psi_{k,i} &= \bm \phi_{k,i-1} - \mu_k \widehat{\nabla_{w^{\sf T}}J}_k(\bm \phi_{k,i-1})\label{equ2a}\\
\bm w_{k,i} &= \displaystyle{\sum_{\ell\in\mathcal{N}_k}}a_{2,\ell k}\bm \psi_{\ell,i}\label{equ2b}
\end{empheq}
\end{subequations}
where the $M$-vector $\bm w_{k,i}$ denotes the estimate by agent $k$ at iteration $i$ for $w^o$, while $\bm{\psi}_{k,i}$ and $\bm \phi_{k,i-1}$ are intermediate estimates. Moreover, an approximation for the true gradient vector of $J_k(w)$, $\widehat{\nabla_{w^{\sf T}}J}_k(\cdot)$, is used in (\ref{equ2a}) since it is generally the case that the true gradient vector is not available (e.g., when $J_k(w)$ is defined as the expectation of some loss function and the probability distribution of the data is not known beforehand to enable computation of $J_k(\cdot)$ or its gradient vector). The symbol ${\cal N}_k$ in (\ref{equ2}) refers to the neighborhood of agent $k$. The $N\times N$ combination matrices $A_1=[a_{1,\ell k}]$ and $A_2=[a_{2,\ell k}]$ are left-stochastic matrices consisting of convex combination coefficients that satisfy:
\begin{equation}\label{equ3}
a_{j,\ell k}\geq0,\;\;\sum\limits_{\ell=1}^{N}a_{j,\ell k}=1,\;\;a_{j,\ell k}=0,\textrm{if}\,\, \ell\notin \mathcal{N}_{k}
\end{equation}
for $j =1,2$. Either of these two matrices can be chosen as the identity matrix, in which case algorithm (\ref{equ2}) reduces
to one of two common forms for diffusion adaptation: the adapt-then-combine (ATC) form when $A_1=I$
and the combine-then-adapt (CTA) form when $A_2=I$. We continue to work with the general
formulation (\ref{equ2}) in order to treat both algorithms, and other cases as well, in a unified manner. The parameter $\mu_k>0$ is a constant step-size factor used to drive the adaptation process. Its value is set to a constant in order to enable continuous adaptation in response to streaming data or drifting minimizers. We could also consider a distributed implementation of the useful consensus-type \cite{ASayed2014, Moura2012,Nedic2010,Kar2011,Dimakis2010,Sardellitti2010,Braca2008,Khan2008,Xiao2004}. However, it has been shown in \cite{ASayed2014, Tu12} that when constant step-sizes are used to drive adaptation, the diffusion networks have wider stability ranges and superior  performance. This is because consensus implementations have an inherent asymmetry in their updates, which can cause network graphs to behave in an unstable manner when the step-size is  constant. This problem does not occur over diffusion networks. Since adaptation is a core element of the proposed strategies in this work, we therefore focus on diffusion learning mechanisms.

The main distinction in this work relative to prior studies on diffusion or consensus adaptive networks is that we now assume that, at each iteration $i$, the adaptation step in (\ref{equ2a}) has only access to a {\em random subset} of the entries of the approximate gradient vector. This situation may arise due to missing data or a purposeful desire to reduce the computational burden of the update step. We model this scenario by replacing the approximate gradient vector by
\begin{equation}\label{equ4}
\widehat{\nabla_{w^{\sf T}}J}^{\rm miss}_k(\bm \phi_{k,i-1}) = \bm{\Gamma}_{k,i}\cdot\widehat{\nabla_{w^{\sf T}}J}_k(\bm \phi_{k,i-1})
\end{equation}
where the random matrix $\bm{\Gamma}_{k,i}$ is diagonal and consists of Bernoulli random variables $\{\bm{r}_{k,i}(m)\}$; each of these variables is either zero or one with probability
\begin{equation}\label{equ4a}
\mbox{\rm Prob}(\bm{r}_{k,i}(m)=0)\triangleq r_k
\end{equation}
where $0\leq r_k<1$ and
\begin{equation}\label{equ5}
\bm{\Gamma}_{k,i} = \mbox{\rm diag}\{\bm{r}_{k,i}(1),\bm{r}_{k,i}(2),\ldots,\bm{r}_{k,i}(M)\}
\end{equation}
In the case when $\bm{r}_{k,i}(m)=0$, the $m$-th entry of the gradient vector is missing, and then the $m$-th entry of $\bm \psi_{k,i}$ in (\ref{equ2a}) is not updated. Observe that we are attaching two subscripts to $\bm{r}$: $k$ and $i$, which means that we are allowing the randomness in the update to vary across agents and also over time.

\subsection{Relation to Block-Coordinate Descent Methods}
Our formulation provides a nontrivial generalization of the powerful random coordinate-descent
technique studied, for example, in the context of deterministic optimization in \cite{nesterov2012,richtarik2014,lu2013} and the references
therein. Random coordinate-descent has been primarily applied in the literature to single-agent
convex optimization, namely, to problems of the form:
\begin{equation}
w^o=\arg\min_w J(w)
\end{equation}
where $J(w)$ is assumed to be known beforehand. The traditional gradient descent algorithm for
seeking the minimizer of $J(w)$, assumed differentiable, takes the form
\begin{equation}
w_i=w_{i-1}-\mu \nabla_{w^{\sf T}} J(w_{i-1})
\end{equation}
where the full gradient vector is used at every iteration to update $w_{i-1}$ to $w_i$. In a coordinate-descent implementation,
on the other hand, at every iteration $i$, only a subset of the entries of the gradient vector is
used to perform the update. These subsets are usually chosen as follows. First, a collection of $K$ partitions of
the parameter space $w$ is defined. These partitions are defined by diagonal matrices, $\{\Omega_k\}$.
Each matrix has ones and zeros on the diagonal and the matrices add up to the identity matrix:
\begin{equation}
\sum_{k=1}^K \Omega_k=I_M
\end{equation}
Multiplying $w$ by any $\Omega_k$ results in a vector of similar size, albeit one where the only
nontrivial entries are those extracted by the unit locations in $\Omega_k$. At every iteration $i$, one of
the partitions is selected randomly, say, with probability
\begin{equation}
\mbox{\rm Prob}(\bm{\Gamma}_i=\Omega_k)=\omega_k
\end{equation}
where the $\{\omega_k\}$ add up to one. Subsequently, the gradient descent iteration is replaced by
\begin{equation}\label{equ5a}
\bm w_i=\bm w_{i-1}-\mu \bm{\Gamma}_i \nabla_{w^{\sf T}} J(\bm w_{i-1})
\end{equation}
This formulation is known as the randomized block-coordinate descent (RBCD) algorithm \cite{nesterov2012,richtarik2014,lu2013}. At each
iteration, the gradient descent step employs only a collection of coordinates represented by the selected
entries from the gradient vector. Besides reducing complexity, this step helps alleviate the condition on the step-size parameter for convergence.

If we reduce our formulation (\ref{equ2})--(\ref{equ4}) to the single agent case, it will become similar to (\ref{equ5a}) in that the desired cost function is optimized only along a \emph{subset} of the coordinates at each iteration. However, our algorithm offers more randomness in generating the coordinate blocks than the RBCD algorithm, by allowing more random combinations of the coordinates at each time index. In particular, we do not limit the selection of the coordinates to a collection of $K$ possibilities
pre-determined by the $\{\Omega_k\}$. Moreover, in our work we use a random subset of the \emph{{stochastic}} gradient vector instead of the \emph{true} gradient vector to update the estimate, which is necessary for adaptation and online learning when the true risk function itself is not known (since the
statistical distribution of the data is not known). Also, our results consider a general multi-agent
scenario involving distributed optimization where \emph{each} individual agent employs random coordinates for
its own gradient direction, and these coordinates are generally different from the coordinates used by other agents. In other
words, the networked scenario adds significant flexibility into the operation of the agents under model
(\ref{equ4}).
\subsection{Relation to Partial Updating Schemes}
It is also important to clarify the differences between our formulation and other works in the
literature, which rely on other useful notions of partial information updates. To begin with, our formulation (\ref{equ4}) is different from the models used in \cite{Zhao20151,Zhao20152,Zhao20153} where the step-size parameter was modeled as a random Bernoulli variable, $\bm{\mu}_k(i)$, which could assume the values $\mu_k$ or zero with certain probability. In that case, when the step-size is zero, all entries of $\bm \psi_{k,i}$ will not be updated and adaptation is turned off completely. This is in contrast to the current scenario where only a subset of the entries are left without update and, moreover, this subset varies randomly from one iteration to another.

Likewise, the useful work \cite{Arablouei2014} employs a different notion of partial sharing of information by focusing on
the exchange of partial entries of the weight estimates themselves rather than on partial entries of the
gradient directions. In other words, the partial information used in this work relates to the combination steps (\ref{equ2c}) and (\ref{equ2b}) rather than to the adaptation step (\ref{equ2a}). It also focuses on the special case in which the risks $\{J_k(w)\}$ are quadratic in $w$. In \cite{Arablouei2014}, it is assumed that only a subset of the weight entries are shared (diffused) among neighbors and that
the estimate itself is still updated fully. In comparison, the formulation we are considering diffuses all entries of the weight estimates. Similarly, in \cite{Gholami2013} it is assumed that some entries of the regression vector are missing, which causes changes to the gradient vectors. In order to undo these changes, an estimation scheme is proposed in \cite{Gholami2013} to estimate the missing data. In our formulation, more generally, a random subset of the entries of the gradient vector are set to zero at each iteration, while the remaining entries remain unchanged and do not need to be estimated.

There are also other criteria that have been used in the literature to motivate partial updating. For example, in \cite{Douglas1997}, the periodic and sequential least-mean-squares (LMS) algorithms are proposed, where the former scheme updates the whole coefficient vector every $N-$th iteration, with $N>1$, and the latter updates only a fraction of the coefficients, which are pre-determined, at each iteration. In \cite{Werner2008,Werner2010ASILOMAR} the weight vectors are partially updated by following a set-membership approach, where updates occur only when the {\em innovation} obtained from the data exceeds a predetermined threshold. In \cite{Douglas1994,Dogancay2001,Werner2010ASILOMAR}, only entries corresponding to the largest magnitudes in the regression vector  or the gradient vector at each agent are updated. However, such scheduled updating techniques may suffer from non-convergence in the presence of nonstationary signals \cite{Godavarti2005}. Partial update schemes can also be based on dimensionality reduction policies using Krylov subspace concepts \cite{Chouvardas2013,Theodoridis2011,Chouvardas2012}. There are also techniques that rely on energy considerations to limit updates, e.g., \cite{Gharehshiran2013}.

The objective of the analysis that follows is to examine the effect of {\em random} partial gradient information on the learning performance and convergence rate of adaptive networks for general risk functions. We clarify these questions by adapting the framework described in \cite{ASayed2014,Sayed2014}.

{\emph{Notation}}: We use lowercase letters to denote vectors, uppercase
letters for matrices, plain letters for deterministic
variables, and boldface letters for random variables. We also
use $(\cdot)^{\sf T}$ to denote transposition, $(\cdot)^{-1}$ for matrix inversion,
$\mathrm{Tr}(\cdot)$ for the trace of a matrix, $\diag\{\cdot\}$ for a diagonal matrix, $\mathrm{col}\{\cdot\}$ for a column vector, $\lambda(\cdot)$ for the eigenvalues of
a matrix, $\rho(\cdot)$ for the spectral radius of a matrix, $\|\cdot\|$ for the two-induced norm of a matrix or the Euclidean
norm of a vector, $\|x\|_{\Sigma}^2$ for the weighted square value $x^{\sf T}\Sigma x$, $\otimes$ for Kronecker product, $\otimes_b$ for block Kronecker product.
Besides, we use $p\succ 0$ to denote that all entries of vector $p$ are positive. Moreover, $\alpha=O(\mu)$ signifies that $|\alpha|\leq c|\mu|$ for some constant $c > 0$, and $\alpha=o(\mu)$ signifies that $\alpha/\mu\to0$ as $\mu\to0$. In addition, the notation $\limsup_{n\to\infty}a(n)$ denotes limit superior of the sequence $a(n)$.

\section{Data Model and Assumptions}
Let $\bm{\mathcal{F}}_{i-1}$ represent the filtration (collection) of all random events generated by the processes $\{\bm w_{k,j}\}$ and $\{\bm\Gamma_{k,j}\}$ at all agents up to time $i-1$. In effect, the notation $\bm{\mathcal F}_{i-1}$ refers to the collection of all past $\{\w_{k,j},\bm{\Gamma}_{k,j}\}$ for all $j\leq i-1$ and all agents.

\begin{assumption}\label{ass2}
(\textbf{Conditions on indicator variables}). It is assumed that the indicator variables $\bm{r}_{k,i}(m)$ and $\bm{r}_{\ell,i}(n)$ are independent of each other, for all $\ell,k,m,n$. In addition, the variables $\{\bm{r}_{k,i}(m)\}$ are independent of $\bm{\mathcal{F}}_{i-1}$ and $\widehat{\nabla_{w^{\sf T}}J}_k(\bm w)$ for any iterates $\bm w\in\bm{\mathcal{F}}_{i-1}$ and for all agents $k$.
\hfill $\Box$
\end{assumption}

Let
\begin{equation}\label{equ16}
\bm s_{k,i}(\bm{\phi}_{k,i-1}) \triangleq \widehat{\nabla_{w^{\sf T}}J}_k(\bm{\phi}_{k,i-1})-{\nabla_{w^{\sf T}}J}_k(\bm{\phi}_{k,i-1})
\end{equation}
denote the gradient noise at agent $k$ at iteration $i$, based on the {\em complete} approximate gradient vector, $\widehat{\nabla_{w^{\sf T}}J}_k(\bm w)$.
We introduce its conditional second-order moment
\begin{equation}\label{equ41}
\bm R_{s,k,i}(\bm w)\triangleq\mathbb{E}[\bm s_{k,i}(\bm w)\bm s^{\sf T}_{k,i}(\bm w)|\bm{\mathcal{F}}_{i-1}].
\end{equation}
The following assumptions are standard and are satisfied by important cases of interest, such as logistic regression risks or mean-square-error risks, as already shown in \cite{ASayed2014,Sayed2014}. These references also motivate these conditions and explain why they are reasonable.

\begin{assumption}\label{ass4}(\textbf{Conditions on gradient noise}) \cite[pp. 496--497]{ASayed2014}. It is assumed that the first and
fourth-order conditional moments of the individual gradient noise processes satisfy the following conditions for any iterates
$\bm w \in  \bm{\mathcal{F}}_{i-1} $ and for all $k,\ell= 1, 2,\ldots,N$:
\begin{align}
\mathbb{E}[\bm s_{k,i}(\bm w)|\bm{\mathcal{F}}_{i-1}]&=0\label{equ20}\\
\mathbb{E}[\bm s_{k,i}(\bm w)\bm s_{\ell,i}^{\sf{T}}(\bm w)|\bm{\mathcal{F}}_{i-1}]&=0,\,k\neq \ell\label{equ22}\\
\mathbb{E}[\|\bm s_{k,i}(\bm w)\|^4|\bm{\mathcal{{F}}}_{i-1}]&\leq{\beta}_k^4\|\bm w\|^4+{\sigma}_{s,k}^4 \label{equ23a}
\end{align}
almost surely, for some nonnegative scalars ${\beta}_k^4$ and ${\sigma}_{s,k}^4$.\hfill $\Box$
\end{assumption}
\begin{assumption}\label{ass5} (\textbf{Smoothness conditions}) \cite[pp. 552,576]{ASayed2014}.
It is assumed that the Hessian matrix of each individual cost function, $J_k(w)$,
and the covariance matrix of each individual gradient noise process are locally Lipschitz continuous in a small neighborhood around $w = w^o$ in the following manner:
\begin{align}
\|\nabla_w^2J_k(w^o+\triangle w)-\nabla_w^2J_k(w^o)\|&\leq \kappa_c\|\triangle w\|\label{equ7a}\\
\left\|\bm R_{s,k,i}(w^o+\triangle w)-\bm R_{s,k,i}(w^o)\right\|&\leq \kappa_d\|\triangle w\|^\gamma\label{equ7b}
\end{align}
for any small perturbations $\|\triangle w\|\leq\varepsilon$ and for some $\kappa_c\geq0$, $\kappa_d\geq0$, and parameter $0<\gamma\leq4$.\hfill $\Box$
\end{assumption}

\section{Main Results: Stability and Performance}
For each agent $k$, we introduce the error vectors:
\begin{align}
\widetilde{\bm{{w}}}_{k,i}&\triangleq w^o-\bm w_{k,i}\label{equ12}\\
\widetilde{\bm{{\phi}}}_{k,i}&\triangleq w^o-\bm \phi_{k,i}\label{equ12b}\\
\widetilde{\bm{{\psi}}}_{k,i}&\triangleq w^o-\bm \psi_{k,i}\label{equ13}
\end{align}
We also collect all errors, along with the gradient noise processes, from across the network into block vectors:
\begin{align}
\widetilde{\bm{{w}}}_{i} &\triangleq \mathrm{col}\left\{\widetilde{\bm{{w}}}_{1,i},\widetilde{\bm{{w}}}_{2,i},\ldots,\widetilde{\bm{{w}}}_{N,i}\right\}\label{equ14a}\\
\widetilde{\bm{{\psi}}}_{i}&\triangleq \mathrm{col}\left\{\widetilde{\bm{{\psi}}}_{1,i},\widetilde{\bm{{\psi}}}_{2,i},\ldots,\widetilde{\bm{{\psi}}}_{N,i}\right\}\label{equ14b}\\
\widetilde{\bm{{\phi}}}_{i} &\triangleq \mathrm{col}\left\{\widetilde{\bm{{\phi}}}_{1,i},\widetilde{\bm{{\phi}}}_{2,i},\ldots,\widetilde{\bm{{\phi}}}_{N,i}\right\}\label{equ14c}\\
{\bm{{s}}}_{i} &\triangleq \mathrm{col}\left\{{\bm{{s}}}_{1,i},{\bm{{s}}}_{2,i},\ldots,{\bm{{s}}}_{N,i}\right\}\label{equ14d}
\end{align}
For simplicity, in (\ref{equ14d}) we use the notation $\bm s_{k,i}$ to replace the gradient noise $\bm s_{k,i}(\bm \phi_{k,i-1})$ defined in (\ref{equ16}), but note vector ${\bm{{s}}}_{i}$ is dependent on the collection of $\{\bm \phi_{k,i-1}\}$ for all $k$.
We further introduce the extended matrices:
\begin{align}
\mathcal{M}&\triangleq\mbox{\rm diag}\{\mu_1,\mu_2,\ldots,\mu_N\}\otimes I_{M}\label{equ160}\\
\mathcal{A}_1&\triangleq A_1\otimes I_{M},\mathcal{A}_2\triangleq A_2\otimes I_{M}\label{equ160b}\\
\bm{\Gamma}_i&\triangleq\mbox{\rm diag}\{{\bm{{\Gamma}}}_{1,i},{\bm{{\Gamma}}}_{2,i},\ldots,{\bm{{\Gamma}}}_{N,i}\}\label{equ160c}
\end{align}
Note that the main difference between the current work and the prior work in \cite{ASayed2014} is the appearance of the random matrices $\{\bm \Gamma_{k,i}\}$ defined by (\ref{equ4}). In the special case when the random matrices are set to the identity matrices  across the agents, i.e., $\{\bm \Gamma_{k,i}\equiv I_M\}$, current coordinate-descent case will reduce to the full-gradient update studied in \cite{ASayed2014}. The inclusion of the random matrices $\{\bm{\Gamma}_{k,i}\}$ adds a non-trivial level of complication because now, agents update only random entries of their iterates at each iteration and, importantly, these entries vary randomly across the agents. This procedure adds a rich level of randomness into the operation of the multi-agent system.   As the presentation will reveal, the study of the stability and limiting performance under these conditions is more challenging than in the stochastic full-gradient diffusion implementation due to at least two factors: (a) First, the evolution of the error dynamics will now involve a {\em non-symmetric} matrix (matrix $\bm{D}_{11,i}$ defined later in (\ref{app9})); because of this asymmetry, the arguments of \cite{ASayed2014} do not apply and need to be modified; and (b) second, there is also randomness in the coefficient matrix for the error dynamics (namely, randomness in the matrix $\bm{\mathcal B}_{i}'$ defined by (\ref{equ31})). These two factors add nontrivial complications to the stability, convergence, and performance analysis of distributed  coordinate-descent solutions, as illustrated by the extended derivations in Appendices \ref{APP1} and \ref{APP3}. These derivations illustrate the new arguments  that are necessary to handle the networked solution of this manuscript. For this reason, in the presentation that follows, whenever we can appeal to a result from \cite{ASayed2014}, we will simply refer to it so that, due to space limitations, we can focus the presentation on the new arguments and proofs that are necessary for the current context. It is clear from the proofs in Appendices \ref{APP1} and \ref{APP3} that these newer arguments are demanding and not straightforward.

\begin{lemma}\label{lem1}
(\textbf{Network error dynamics}). Consider a network of $N$ interacting
agents running the diffusion strategy (\ref{equ2}) with the gradient vector replaced by (\ref{equ4}). The evolution of the error
dynamics across the network relative to the reference vector $w^o$ is described by the following recursion:
\begin{equation}\label{equ24}
\widetilde{\bm{w}}_i = \bm{\mathcal{B}}_{i}\widetilde{\bm{w}}_{i-1}+\mathcal{A}_2^{\sf{T}}\mathcal{M}\bm{\Gamma}_i\bm{s}_i
\end{equation}
where
\begin{align}
\bm{\mathcal{B}}_{i}&\triangleq\mathcal{A}_2^{\sf{T}}\left(I-\mathcal{M}\bm{\Gamma}_i\bm{\mathcal{H}}_{i-1}\right)\mathcal{A}_1^{\sf{T}}\label{equ25}\\
\bm{\mathcal{H}}_{i-1}&\triangleq\mbox{\rm diag}\{\bm H_{1,i-1},\,\bm H_{2,i-1},\ldots,\,\bm H_{N,i-1}\}\label{equ26}\\
\bm H_{k,i-1}&\triangleq\int_0^1\nabla^2_wJ_k(w^o-t\widetilde{\bm\phi}_{k,i-1})dt\label{equ27}.
\end{align}
\end{lemma}

\noindent\emph {Proof}: Refer to \cite[pp. 498--504]{ASayed2014}, which is still applicable to the current context. We only need to set in that derivation the matrix $A_o$ to $A_o =I$, and the vector $b$ to $b=0_{MN}$. These quantities were defined in (8.131) and (8.136) of \cite{ASayed2014}. The same derivation will lead to (\ref{equ24})--(\ref{equ27}), with the main difference being the appearance now of the random matrix $\bm{\Gamma}_i$ in (\ref{equ24}) and (\ref{equ25}). \hfill $\Box$

We assume that the matrix product $P=A_1A_2$ is primitive. This condition is guaranteed automatically,
for example, for ATC and CTA scenarios when the network is strongly-connected. This means, in view of the Perron-Frobenius Theorem \cite{ASayed2014,Sayed2014}, that $P$ has a single eigenvalue at one. We denote the corresponding eigenvector by $p$, and normalize the entries of $p$ to add up to one. It follows from the same theorem that the entries of $p$ are strictly positive, written as
\begin{equation}\label{equ11a}
Pp=p,\;\;\mathds{1}^{\sf T} p=1,\;\;p\succ0
\end{equation}
with $\mathds{1}$ being the vector of size $N$ with all its entries equal to one.
\begin{theorem}\label{theo1}
(\textbf{Network stability}). Consider a strongly-connected network of $N$ interacting agents running the diffusion strategy (\ref{equ2}) with the gradient vector replaced by (\ref{equ4}). Assume the matrix product $P=A_1A_2$ is primitive. Assume also that the individual cost functions, $J_k(w)$, satisfy the condition in (\ref{equ6}) and that Assumptions \ref{ass2}--\ref{ass4} hold. Then, the second and fourth-order moments of the network error vectors are stable for sufficiently small step-sizes, namely, it holds, for all $k= 1, 2,\ldots,N$, that
\begin{align}
\limsup_{i\to\infty}\mathbb{E}\|{\widetilde{\bm{w}}}_{k,i}\|^2&=O(\mu_{\mathrm{max}})\label{equ28a}\\
\limsup_{i\to\infty}\mathbb{E}\|{\widetilde{\bm{w}}}_{k,i}\|^4&=O(\mu^2_{\mathrm{max}})\label{equ28b}
\end{align}
for any $\mu_{\mathrm{max}}<\mu_o$, for some small enough $\mu_o$, where
\begin{equation}\label{app13}
\mu_{\mathrm{max}}\triangleq\max\{\mu_1,\mu_2,\ldots,\mu_N\}.
\end{equation}
\end{theorem}
\noindent\emph {Proof}:
The argument requires some effort and is given in Appendix \ref{APP1}.\hfill $\Box$

\begin{lemma}\label{lem2} (\textbf{Long-term network dynamics}). Consider a strongly-connected network of $N$ interacting agents running the diffusion strategy (\ref{equ2}) under (\ref{equ4}). Assume the matrix product $P=A_1A_2$ is primitive. Assume also that the individual cost functions satisfy (\ref{equ6}), and that Assumptions \ref{ass2}--\ref{ass4} and (\ref{equ7a}) hold. After sufficient
iterations, $i\gg1$, the error dynamics of the network relative to the reference vector $w^o$ 
is well-approximated by the following model:
\begin{equation}\label{equ30}
\widetilde{\bm{w}}_i' = \bm{\mathcal{B}}'_{i}\widetilde{\bm{w}}_{i-1}'+\mathcal{A}_2^{\sf{T}}\mathcal{M}\bm{\Gamma}_i\bm{s}_i,\,\, i\gg 1
\end{equation}
where
\begin{align}
\bm{\mathcal{B}}_{i}'&\triangleq\mathcal{A}_2^{\sf{T}}\left(I-\mathcal{M}\bm{\Gamma}_i\mathcal{H}\right)\mathcal{A}_1^{\sf{T}}\label{equ31}\\
{\mathcal{H}}&\triangleq\mbox{\rm diag}\{ H_{1},\,H_{2},\ldots,\,H_{N}\}\label{equ32}\\
H_{k}&\triangleq\nabla^2_wJ_k(w^o)\label{equ33}
\end{align}
More specifically, it holds for sufficiently small step-sizes that
\begin{align}\label{equ35}
\limsup_{i\to\infty}\mathbb{E}\|{\widetilde{\bm{w}}}_{k,i}'\|^2&=O(\mu_{\mathrm{max}})\\
\limsup_{i\to\infty}\mathbb{E}\|{\widetilde{\bm{w}}}_{k,i}'\|^4&=O(\mu_{\mathrm{max}}^2)\label{equ35b}
\end{align}
\begin{equation}\label{equ34}
\limsup_{i\to\infty}\mathbb{E}\|\widetilde{\bm{w}}_i'\|^2=\limsup_{i\to\infty}\mathbb{E}\|\widetilde{\bm{w}}_i\|^2+O(\mu_{\mathrm{max}}^{3/2}).
\end{equation}
\end{lemma}
\noindent\emph {Proof}:\label{pr2}
To establish (\ref{equ30}), we refer to the derivation in \cite[pp. 553--555]{ASayed2014}, and note that, in our case, $\|\bm{\Gamma}_i\|\leq 1$ and $b={0}_{MN}$ (which appeared in (10.2) of \cite{ASayed2014}). Moreover, the results in (\ref{equ35}) and (\ref{equ35b}) can be established by following similar techniques to the proof of Theorem \ref{theo1}, where the only difference is that the random matrix $\bm{\mathcal{H}}_{i-1}$ defined in (\ref{equ26}) is now replaced with the deterministic matrix $\mathcal{H}$ defined by (\ref{equ32}), and by noting that the matrices $\{H_k\}$ in (\ref{equ33}) still satisfy the condition (\ref{equ6}). With regards to result (\ref{equ34}), we refer to the argument in \cite[pp. 557--560]{ASayed2014} and note again that $\|\bm{\Gamma}_i\|\leq 1$.\hfill$\Box$

Result (\ref{equ28a}) ensures that the mean-square-error (MSE) performance of the network is in the order of $\mu_{\max}$. Using the long-term model (\ref{equ30}), we can be more explicit and derive the proportionality constant that describes the value of the network mean-square-error to first-order in $\mu_{\max}$. To do so, we introduce the quantity
\begin{equation}\label{equ11}
q\triangleq \mbox{\rm diag}\left\{\mu_1,\mu_2,\ldots,\mu_N\right\}A_2 p
\end{equation}
and the gradient-noise covariance matrices:
\begin{align}
G_k&\triangleq\lim_{i\to\infty}\bm R_{s,k,i}(w^o)\label{app45e}\\
G_k'&\triangleq\mathbb{E}[\bm\Gamma_{k,i}G_k\bm\Gamma_{k,i}]\label{equ45d}.
\end{align}
Observe that $G_k$ is the limiting covariance matrix of the gradient noise process evaluated at $w^o$, and is assumed to be a constant value, while $G_k'$ is a
weighted version of it. A typical example for the existence of the limit in (\ref{app45e}) is the MSE network, where the covariance matrix of the gradient noise is a constant matrix, which is independent of the time index $i$ \cite[p. 372]{ASayed2014}. It follows by direct inspection that the entries of $G_k'$ are given by:
\begin{equation}\label{equ46}
G_k'(m,n)=\left\{\begin{array}{ll}(1-r_k)^2G_k(m,n),&m\neq n\\
(1-r_k)G_k(m,m),&m=n.\end{array}\right.
\end{equation}
We also define the mean-square-deviation (MSD) for each agent $k$, and the average MSD across the network to first-order in $\mu_{\max}$ --- see \cite{ASayed2014} for further clarifications on these expressions where it is explained, for example, that $\mbox{\rm MSD}_k$ provides the steady-state value of the error variance $\mathbb{E}\|
\widetilde{\bm w}_{k,i}\|^2$ to first-order in $\mu_{\max}$:
\begin{align}
\mathrm{MSD}_k &\triangleq\mu_{\mathrm{max}}\left(\lim_{\mu_\mathrm{max}\to0}\limsup_{i\to\infty}\frac{1}{\mu_{\mathrm{max}}}\mathbb{E}\|\bm{\widetilde{w}}_{k,i}\|^2\right)\label{equ42}\\
\mathrm{MSD}_{av} &\triangleq\frac{1}{N}\sum\limits_{k=1}^N\mathrm{MSD}_k \label{equ42b}.
\end{align}
Likewise, we define the excess-risk (ER) for each agent $k$ as the average fluctuation of the normalized aggregate cost
\begin{equation}\label{equ68}
\bar{J}^{\rm glob}(w)\triangleq\left(\sum\limits_{k=1}^{N}q_k\right)^{-1}\sum\limits_{k=1}^{N}q_kJ_k(w)
\end{equation}
with $\{q_k\}$ being entries of the vector $q$ defined by (\ref{equ11}), around its minimum value $\bar{J}^{\rm glob}(w^o)$ at steady state to first-order in $\mu_{\max}$, namely \cite[p. 581]{ASayed2014}:
\begin{multline}\label{equ69}
\mathrm{ER}_k \triangleq\mu_{\mathrm{max}}\\\times\left(\lim_{\mu_\mathrm{max}\to0}\limsup_{i\to\infty}\frac{1}{\mu_{\mathrm{max}}}\mathbb{E}\{\bar{J}^{\rm glob}(\bm{{w}}_{k,i})-\bar{J}^{\rm glob}({{w}}^o)\}\right).
\end{multline}
The average ER across the network is defined by
\begin{equation}\label{equ71}
\mathrm{ER}_{av} \triangleq\frac{1}{N}\sum_{k=1}^N\mathrm{ER}_k.
\end{equation}
By following similar arguments to \cite[p. 582]{ASayed2014}, it can be verified that the excess risk can also be evaluated by computing a weighted mean-square-error
variance:
\begin{equation}
\mathrm{ER}_k \triangleq\mu_{\mathrm{max}}\left(\lim_{\mu_\mathrm{max}\to0}\limsup_{i\to\infty}\frac{1}{\mu_{\mathrm{max}}}\mathbb{E}\|\bm{\widetilde{w}}_{k,i}\|^2_{\frac{1}{2}\bar{H}}\right)\label{equ42c}
\end{equation}
where $\bar{H}$ denotes the Hessian matrix of the normalized aggregate cost, $\bar{J}^{\rm glob}(w)$, evaluated at the minimizer $w=w^o$:
\begin{equation}\label{equ67}
\bar{H}\triangleq\left(\sum\limits_{k=1}^{N}q_k\right)^{-1}\sum_{k=1}^Nq_kH_k
\end{equation}
with $H_k$ defined by (\ref{equ33}).
Moreover, we define the convergence rate as the slowest rate at which the error variances, $\mathbb{E}\|\bm{\widetilde{w}}_{k,i}\|^2$, converge to the steady-state region. By iterating the recursion for the second-order moment of the error vector, we will arrive at a relation in the following form:
\begin{equation}\label{equ94}
\mathbb{E}[\|\widetilde{\bm{w}}_i\|^2]= \mathbb{E}\left\{\|\widetilde{\bm{w}}_{-1}\|^2_{{F}^{i+1}}\right\}+c
\end{equation}
for some matrix ${F}$ and constant $c$, where $\widetilde{\bm{w}}_{-1}$ denotes the network error vector at the initial time instant. The first-term on the right-hand side corresponds to a transient component
that dies out with time, and the second-term denotes the steady-state region that $\mathbb{E}[\|\widetilde{\bm{w}}_i\|^2]$ converges to. Then, the convergence rate of $\mathbb{E}[\|\widetilde{\bm{w}}_i\|^2]$ towards its steady-state region is dictated by $\rho({F})$ \cite[p. 395]{ASayed2014}. The following conclusion is one of the main results in this work. It shows how the coordinate
descent construction influences performance in comparison to the standard diffusion strategy where all
entries of the gradient vector are used at each iteration. Following the statement of the result, we
illustrate its implications by considering several important cases.
\begin{theorem}\label{theo2} (\textbf{MSD and ER performance}). Under the same setting of Theorem \ref{theo1}, and assume also that Assumption \ref{ass5} holds, it holds that, for sufficiently small step-sizes:
\begin{multline}\label{equ43}
\mathrm{MSD}_{\mathrm{coor},k}= \mathrm{MSD}_{\mathrm{coor,av}}\\=\frac{1}{2}\mathrm{Tr}\left(\left(\sum\limits_{k=1}^Nq_k(1-r_k)H_k\right)^{-1}\sum\limits_{k=1}^Nq_k^2G_k'\right)
\end{multline}
\begin{equation}\label{equ72}
\mathrm{ER}_{\mathrm{coor},k}= \mathrm{ER}_{\mathrm{coor,av}}=\frac{1}{2}\mathrm{Tr}\left(X\sum\limits_{k=1}^Nq_k^2G_k'\right)
\end{equation}
where the subscript ``$\mathrm{coor}$'' denotes the stochastic coordinate-descent diffusion implementation, and matrix $X$ is the unique solution to the following Lyapunov equation:
\begin{equation}\label{equ73}
X\left(\sum\limits_{k=1}^Nq_k(1-r_k)H_k\right)+ \left(\sum\limits_{k=1}^Nq_k(1-r_k)H_k\right)X=\bar{H}
\end{equation}
with $\bar{H}$ defined by (\ref{equ67}). Moreover, for large enough $i$, the convergence rate of the error variances, $\mathbb{E}\|\bm{\widetilde{w}}_{k,i}\|^2$,
towards the steady-state region (\ref{equ43}) is given by
\be\label{equ88}
\alpha_{\mathrm{coor}} = 1-2\lambda_{\min}\left(\sum_{k=1}^Nq_k(1-r_k)H_k\right) + O\left(\mu_{\max}^{(N+1)/N}\right)
\ee
\end{theorem}

\noindent \emph {Proof}: See Appendix \ref{APP3}.
\hfill $\Box$

\section{Implications and Useful Cases} \label{sec4}
\subsection{Uniform Missing Probabilities}
Consider the case when the missing probabilities are identical across the agents, i.e., $\{r_k\equiv r\}$.
\subsubsection{Convergence time} Consider the ATC or CTA forms of the full-gradient or coordinate-descent diffusion
strategy (\ref{equ2c})--(\ref{equ2b}) and (\ref{equ4}). From (\ref{equ94}), we find that the error variances for the distributed strategies evolve according to a relation of the form:
\begin{equation}\label{app103}
\mathbb{E}\|\widetilde{\bm{w}}_{k,i}\|^2\leq {\alpha^{i+1}}\mathbb{E}\|\widetilde{\bm{w}}_{k,-1}\|^2+c
\end{equation}
for some constant $c>0$, and where the parameter $\alpha$ determines the convergence rate. Its value
is denoted by $\alpha_{\rm grad}$ for the full-gradient implementation and is given by \cite[p. 584]{ASayed2014}:
\be
\alpha_{\mathrm{grad}} = 1-2\lambda_{\min}\left(\sum_{k=1}^Nq_kH_k\right) + o\left(\mu_{\max}\right)
\ee
Likewise, the convergence rate for the coordinate-descent variant is denoted by $\alpha_{\rm coor}$
and is given by expression (\ref{equ88}). It is clear that $\alpha_{\mathrm{coor}}\geq\alpha_{\mathrm{grad}}$
for $0\leq r<1$, so that the coordinate-descent implementation converges at a slower rate as expected (since it only
employs partial gradient information). Thus, let $T_{\mathrm{coor}}$ and $T_{\mathrm{grad}}$ denote the largest number of iterations that are needed for the error variances, $\mathbb{E}\|\widetilde{\bm{w}}_{k,i}\|^2$, to converge to their steady-state regions. The values of $T_{{\rm coor}}$ and $T_{{\rm grad}}$ can be estimated by assessing the number of iterations that it takes for the transient term $\alpha^{i+1}\Ex\|\widetilde{\w}_{k,-1}\|^2$ in (\ref{app103}) to assume a higher-order value in $\mu_{\max}$, i.e., for
\begin{align}
\alpha_{\mathrm{coor}}^{T_{\mathrm{coor}}}\mathbb{E}\|\widetilde{\bm{w}}_{k,-1}\|^2&=d\mu_{\max}^{1+\epsilon}\\
\alpha_{\mathrm{grad}}^{T_{\mathrm{grad}}}\mathbb{E}\|\widetilde{\bm{w}}_{k,-1}\|^2&=d\mu_{\max}^{1+\epsilon}
\end{align}
for some proportionality constant $d$, and small number $\epsilon >0$. Then, it holds that
\begin{align}\label{app111}
\frac{T_{\mathrm{coor}}}{T_{\mathrm{grad}}}
&=\frac{{{\rm ln}\,\alpha_{\mathrm{grad}}}}{{{\rm ln}\,\alpha_{\mathrm{coor}}}}\nn
&\stackrel{(a)}\approx\frac{{{\rm ln}\left(1-2\lambda_{\min}\left(\sum_{k=1}^Nq_kH_k\right)\right)}}{{{\rm ln}\left(1-2\lambda_{\min}\left(\sum_{k=1}^Nq_k(1-r)H_k\right)\right)}}\nn
&\stackrel{(b)}\approx\frac{-2\lambda_{\min}\left(\sum_{k=1}^Nq_kH_k\right)}{-2(1-r)\lambda_{\min}\left(\sum_{k=1}^Nq_kH_k\right)}\nn
&=\frac{1}{1-r}
\end{align}
where in step (a) we ignored the higher-order term in $\mu_{\max}$, and in (b) we used $\ln(1-x)\approx -x$ as $x\rightarrow 0$.
Expression (\ref{app111}) reveals by how much the convergence time is increased in the coordinate-descent
implementation. Note that because of longer convergence time, the stochastic coordinate-descent diffusion implementation may require more quantities to be exchanged across the network compared to the full-gradient case.
\subsubsection{Computational complexity}
Let us now compare the computational complexity of both implementations: the coordinate-descent and
the full-gradient versions. Assume that the computation required to calculate each entry of the gradient vector $\widehat{\nabla_{w^{\sf T}}J}_k(\bm \phi_{k,i-1})$ is identical, and let $c_m\geq0$ and $c_a\geq0$ denote the number of multiplications and additions, respectively, that are needed for each entry of the gradient vector.

Let $n_k\triangleq|\mathcal{N}_k|$ denote the degree of
agent $k$. Then, in the full-gradient implementation, the adaptation step (\ref{equ2a}) requires $c_mM+M$ multiplications and $c_aM+M$ additions, while the combination step (\ref{equ2c}) or (\ref{equ2b}) requires $n_kM$ multiplications and $(n_k-1)M$ additions. In the coordinate-descent implementation, the adaptation step (\ref{equ2a}) with the gradient vector replaced by (\ref{equ4}) requires $(1-r)\cdot(c_mM+M)$ multiplications and $(1-r)\cdot(c_aM+M)$ additions on average, while the combination step (\ref{equ2c}) or (\ref{equ2b}) requires $n_kM$ multiplications and $(n_k-1)M$ additions. Let $m_{\mathrm{coor},k}$ and $m_{\mathrm{grad},k}$ denote the combined number of multiplications required by the adaptation and combination steps per iteration at each agent $k$ in the coordinate-descent and full-gradient cases. Then,
\begin{align}
m_{\mathrm{grad},k}&=(c_m+n_k+1)M\label{app108}\\
m_{\mathrm{coor},k}
&=m_{\mathrm{grad},k}-(c_m+1)Mr\label{app106}
\end{align}
If we now consider that these algorithms take $T_{\rm coor}$ and $T_{\rm grad}$ iterations to reach their
steady-state regime, then the total number of multiplications at agent $k$, denoted by $M_{\mathrm{coor},k}$ and $M_{\mathrm{grad},k}$, are therefore given by
\begin{align}
M_{\mathrm{coor},k}&=m_{\mathrm{coor},k}T_{\mathrm{coor}}\label{app104}\\
M_{\mathrm{grad},k}&=m_{\mathrm{grad},k}T_{\mathrm{grad}}\label{app105}
\end{align}
so that using (\ref{app111}):
\be\label{app109}
\frac{M_{\mathrm{coor},k}}{M_{\mathrm{grad},k}}=\frac{m_{\mathrm{coor},k}}{m_{\mathrm{grad},k}}\frac{1}{1-r}
\ee
Now, the first term on the right hand side satisfies
\begin{align}
\frac{m_{\mathrm{coor},k}}{m_{\mathrm{grad},k}}
&=1-\frac{c_m+1}{c_m+n_k+1}r\label{app115}
\end{align}
so that from (\ref{app109}) and (\ref{app115}):
\be
1\leq\frac{M_{\mathrm{coor},k}}{M_{\mathrm{grad},k}}=(1-r)^{-1}\left(1-\frac{c_m+1}{c_m+n_k+1}r\right)
\ee
since $0\leq r<1$. It is clear that when it is costly to compute the gradient entries, i.e., when $c_m \gg n_k$, then $M_{{\rm
coor},k} $ and $M_{{\rm grad},k}$ will be essentially identical. This means that while the coordinate-descent
implementation will take longer to converge, the savings in computation per iteration that it
provides is such that the overall computational complexity until convergence remains largely invariant (it
is not increased). This is a useful conclusion. It means that in situations where computations at each
iteration need to be minimal (e.g., when low end sensors are used), then a coordinate-descent variant is recommended and it will be able to
deliver the same steady-state performance (to first-order in $\mu_{\max}$, see (\ref{equ93}) ahead) with the total computational demand spread over a longer
number of iterations. This also means that the complexity and convergence rate measures, when
normalized by the number of entries that are truly updated at each iteration, remain effectively
invariant. A similar analysis and conclusion holds if we examine the total number of additions (as opposed
to multiplications) that are necessary.
\subsubsection{MSD performance}
The matrix $G_k'$ defined by (\ref{equ46}) can be written as
\begin{align}\label{equ51}
G_k'&= (1-r)^2G_k+\left((1-r)-(1-r)^2\right)\mbox{\rm diag}\{G_k\}\nn
&=(1-r)^2\left(G_k+\frac{r}{1-r}\mbox{\rm diag}\{G_k\}\right)
\end{align}
where the term $\mbox{\rm diag}\{G_k\}$ is a diagonal matrix that consists of the diagonal entries of $G_k$. Then, the MSD expression (\ref{equ43}) gives
\begin{align}
\mathrm{MSD}_{\mathrm{coor},k}\hspace{0.05cm}
&\hspace{-0.15cm}\stackrel{(\ref{equ51})}=\hspace{-0.1cm}\frac{1}{2}(1-r)\mathrm{Tr}\Bigg(\left(\sum\limits_{k=1}^Nq_kH_k\right)^{-1}\times\nonumber\\
&\hspace{2cm}\sum\limits_{k=1}^Nq_k^2\left(G_k+\frac{r}{1-r}\mbox{\rm diag}\{G_k\}\right)\Bigg)\nonumber
\end{align}
\begin{align}
&=\frac{1}{2}\mathrm{Tr}\left(\left(\sum\limits_{k=1}^Nq_kH_k\right)^{-1}\sum\limits_{k=1}^Nq_k^2G_k\right)+\nonumber\\
&\hspace{0.45cm}\frac{r}{2}\mathrm{Tr}\left(\left(\sum\limits_{k=1}^Nq_kH_k\right)^{-1}\sum\limits_{k=1}^Nq_k^2\mbox{\rm diag}\{G_k\}\right)-\nonumber\\
&\hspace{0.45cm}\frac{r}{2}\mathrm{Tr}\left(\left(\sum\limits_{k=1}^Nq_kH_k\right)^{-1}\sum\limits_{k=1}^Nq_k^2G_k\right).\label{equ52a}
\end{align}
By recognizing that the first item in (\ref{equ52a}) is exactly the MSD expression for the stochastic full-gradient diffusion case \cite[p. 594]{ASayed2014}, which is denoted by ``$\mathrm{MSD}_{\mathrm{grad},k}$'', we get
\begin{multline}\label{equ64}
\mathrm{MSD}_{\mathrm{coor},k}-\mathrm{MSD}_{\mathrm{grad},k}\\
=\frac{r}{2}\mathrm{Tr}\Bigg(\left(\sum\limits_{k=1}^Nq_kH_k\right)^{-1}
\sum\limits_{k=1}^Nq_k^2\check{G}_k\Bigg)
\end{multline}
where
\be\label{equ89}
\check{G}_k\triangleq\mbox{\rm diag}\{G_k\}-G_k.
\ee
We show in Appendix \ref{APP6} that the difference in (\ref{equ64}) can be positive or negative, i.e., the MSD performance can be better or worse in the stochastic coordinate-descent case in comparison to the stochastic full-gradient case. Recall from (\ref{equ42}) that the MSD performance is evaluated to first-order in $\mu_{\max}$. Then, the MSD gap in (\ref{equ64}) is to first-order in the step-size
parameter. Observe that the missing probability $r$ on the right hand side of that equation is independent of $\mu_{\max}$. It thus follows that
\be\label{equ90}
\mathrm{Tr}\Bigg(\left(\sum\limits_{k=1}^Nq_kH_k\right)^{-1}
\sum\limits_{k=1}^Nq_k^2\check{G}_k\Bigg)=O(\mu_{\max}).
\ee
\begin{corollary}\label{cor5}(\textbf{Small missing probabilities}). Let $r=O(\mu_{\max}^\varepsilon)$ for a small number $\varepsilon>0$. It holds that
\be\label{equ93}
\mathrm{MSD}_{\mathrm{coor},k}-\mathrm{MSD}_{\mathrm{grad},k}=O(\mu_{\max}^{1+\varepsilon})=o(\mu_{\max}).
\ee
\end{corollary}

\noindent\emph {Proof}: It follows from (\ref{equ64}) and (\ref{equ90}). \hfill $\Box$

We proceed to provide a general upper bound for the difference between $\mathrm{MSD}_{\mathrm{coor},k}$ and $\mathrm{MSD}_{\mathrm{grad},k}$.
\begin{corollary}\label{cor1}(\textbf{Upper bound}). Under the same conditions of Theorem \ref{theo2}, and when the missing probabilities are uniform, namely, $\{r_k\equiv r\}$, it holds that:
\begin{multline}\label{equ52g}
|\mathrm{MSD}_{\mathrm{coor},k}-\mathrm{MSD}_{\mathrm{grad},k}|\leq\\\frac{r}{2}\left({\sum_{k=1}^Nq_k}\right)^{-1}\left(\frac{1}{\nu_d}-\frac{1}{\delta_d}\right)\sum\limits_{k=1}^Nq_k^2\mathrm{Tr}(G_k)
\end{multline}
where the positive numbers $\nu_d\leq\delta_d$ are defined in (\ref{equ6}), and the matrices $\{G_k\}$ are defined by (\ref{app45e}).
Furthermore, when the matrices $\{H_k\}$ or $\{G_k\}$ are diagonal, it follows that
\be\label{equ52h}
\mathrm{MSD}_{\mathrm{coor},k}=\mathrm{MSD}_{\mathrm{grad},k}
\ee
\end{corollary}

\noindent\emph {Proof}: See Appendix \ref{APP4}.
\hfill $\Box$

\begin{corollary}\label{cor4}(\textbf{Uniform step-sizes}). Continuing with the setting of Corollary \ref{cor1} by assuming now that the step-sizes are uniform across all agents and $A_1=I$ or $A_2=I$ (corresponding to either the ATC or CTA formulations). Let $\{p_k\}$ be entries of the vector $p$ defined by (\ref{equ11a}). Then, in view of (\ref{equ11}) and (\ref{equ11a}), $q_k=\mu p_k$ and the $\{p_k\}$ add up to one. In this case, the sum of the $\{q_k\}$ is equal to $\mu$ and expression (\ref{equ52g}) simplifies to
\begin{multline}\label{equ51a}
|\mathrm{MSD}_{\mathrm{coor},k}-\mathrm{MSD}_{\mathrm{grad},k}|\leq\\\frac{r}{2}\mu\left(\frac{1}{\nu_d}-\frac{1}{\delta_d}\right)\sum\limits_{k=1}^Np_k^2\mathrm{Tr}(G_k).
\end{multline}
\end{corollary}
\hfill $\Box$

Consider now MSE networks where the risk function that is associated with each agent $k$ is the mean-square-error:
\begin{equation}\label{equ48}
J_k(w)=\mathbb{E}(\bm d_k(i)-\bm u_{k,i}w)^2
\end{equation}
where $\bm d_k(i)$ denotes the desired signal, and $\bm u_{k,i}$ is a (row) regression vector. In these networks, the data $\{\bm d_k(i),\bm u_{k,i}\}$ are assumed to be related via the linear regression model
\begin{equation}\label{equ49}
\bm d_k(i)=\bm u_{k,i}w^o+\bm v_k(i)
\end{equation}
where $\bm v_k(i)$ is zero-mean white measurement noise with variance $\sigma_{v,k}^2$ and assumed to be independent of all other random variables. The processes $\{\bm d_{k}(i),\bm u_{k,i},\bm v_{k}(i)\}$ are assumed to be jointly wide-sense stationary random processes. Assume also that the regression data $\{\bm u_{k,i}\}$ are zero-mean, and white over time and space with
\be\label{equ49a}
\Ex \bm u_{k,i}^{\sf T}\bm u_{\ell,j}\triangleq R_{u,k}\delta_{k,\ell}\delta_{i,j}
\ee
where $R_{u,k}>0$, and $\delta_{k,\ell}$ denotes the Kronecker delta sequence. Consider the case when the covariance matrices of the regressors are identical across the network, i.e., $\{R_{u,k}\equiv R_u>0\}$. Then, it holds that \cite[p. 598]{ASayed2014}
\be\label{equ49b}
H_k\equiv 2R_{u},\,\,G_k=4\sigma_{v,k}^2R_{u}.
\ee
Substituting into (\ref{equ64}) we have
\begin{align}\label{equ49c}
&\mathrm{MSD}_{\mathrm{coor},k} - \mathrm{MSD}_{\mathrm{grad},k}\nn
&\hspace{0.2cm}=r\left(\sum\limits_{k=1}^Nq_k\right)^{-1}\left({\sum\limits_{k=1}^Nq_k^2\sigma_{v,k}^2}\right)\mathrm{Tr}\left(R_u^{-1}\diag\{R_u\}-M\right)\nn
&\hspace{0.2cm}\geq0
\end{align}
where (\ref{equ49c}) holds because $\mathrm{Tr}\left(R_u^{-1}\diag\{R_u\}\right)\geq M$, which can be shown by using the property that $\mathrm{Tr}\left(X\right)\mathrm{Tr}\left(X^{-1}\right)\geq M^2$ for any $M\times M$ symmetric positive-definite matrix $X$ \cite[p. 317]{ASayed2008}, and choosing $X=\diag^{\frac{1}{2}}\{R_u\}R_u^{-1}\diag^{\frac{1}{2}}\{R_u\}$.
In the case of MSE networks, by exploiting the special relation between the matrices $\{H_{k}\}$ and $\{G_k\}$ in (\ref{equ49b}), we are able to show that the MSD in the stochastic coordinate-descent case is always larger (i.e., worse) than or equal to that in the stochastic full-gradient diffusion case (although by not more than $o(\mu_{\max})$, as indicated by (\ref{equ93})).  We are also able to provide a general upper bound on the difference between these two MSDs.
\begin{corollary}\label{cor2}(\textbf{MSE networks}).
Under the same conditions of Corollary \ref{cor1}, and for MSE networks with uniform covariance matrices, i.e., $\{R_{u,k}\equiv R_u>0\}$, it holds that
\begin{multline}\label{equ65}
0\leq\mathrm{MSD}_{\mathrm{coor},k}-\mathrm{MSD}_{\mathrm{grad},k}\leq\\{r}\left({\sum_{k=1}^Nq_k}\right)^{-1}\left(\sum\limits_{k=1}^Nq_k^2\sigma_{v,k}^2\right)\left(\frac{\delta_d}{\nu_d}-{1}\right)M
\end{multline}
Moreover, it holds that $\mathrm{MSD}_{\mathrm{coor},k}=\mathrm{MSD}_{\mathrm{grad},k}$ if, and only if, $R_u$ is diagonal.
\end{corollary}
\noindent\emph {Proof}: It follows from Corollary \ref{cor1} by using $\mathrm{Tr}\left(G_k\right)=4\sigma_{v,k}^2\mathrm{Tr}\left(R_{u}\right)$ and noting that $\nu_d/2\leq\lambda\left(R_u\right)\leq\delta_d/2$ according to (\ref{equ49b}) and (\ref{equ6}).
\hfill $\Box$
\subsubsection{ER performance}
Consider the scenario when the missing probabilities are identical across the agents, i.e., $\{r_k\equiv r\}$. Then, expression (\ref{equ73}) simplifies to
\begin{equation}\label{equ73b}
(1-r)\left(\sum\limits_{k=1}^Nq_k\right)X\bar{H}+ (1-r)\left(\sum\limits_{k=1}^Nq_k\right)\bar{H}X=\bar{H}
\end{equation}
where we used the equality $\sum_{k=1}^Nq_kH_k=\left(\sum_{k=1}^Nq_k\right)\bar{H}$, it follows that
\begin{equation}\label{equ730a}
X=\frac{1}{2}(1-r)^{-1}\left(\sum\limits_{k=1}^Nq_k\right)^{-1}I_M.
\end{equation}
Thus, the ER expression in (\ref{equ72}) can be rewritten as:
\begin{align}\label{equ730b}
\mathrm{ER}_{\mathrm{coor},k}
&=\frac{1}{4}(1-r)^{-1}\left(\sum\limits_{k=1}^Nq_k\right)^{-1}\mathrm{Tr}\left(\sum\limits_{k=1}^Nq_k^2G'_k\right)\nn
&\stackrel{(a)}=\frac{1}{4}\left(\sum\limits_{k=1}^Nq_k\right)^{-1}\sum\limits_{k=1}^Nq_k^2\mathrm{Tr}\left(G_k\right)
\end{align}
which is exactly the same result for the full gradient case from \cite[p. 608]{ASayed2014}, and where the equality (a) holds because $\mathrm{Tr}\left(G'_k\right)=(1-r)\mathrm{Tr}\left(G_k\right)$ according to the definition in (\ref{equ46}).
\subsection{Uniform Individual Costs}
Consider the case when the individual costs, $J_k(w)$, are identical across the network, namely, \cite[p. 610]{ASayed2014}
\begin{equation}\label{equ75}
J_k(w)\equiv J(w)\triangleq\mathbb{E}Q(w;\bm x_{k,i})
\end{equation}
where $Q(w;\bm x_{k,i})$ denotes the loss function. In this case, it will hold that the matrices $\{H_k,G_k\}$ are uniform across the agents, i.e.,
\begin{equation}
H_k=\nabla^2_wJ(w^o)\equiv H\label{equ76}
\end{equation}
\begin{equation}
G_k=\mathbb{E}{\nabla_{w^{\sf T}}Q(w^o;\bm x_{k,i})}\left[{\nabla_{w^{\sf T}}Q(w^o;\bm x_{k,i})}\right]^{\sf T}\equiv G
\end{equation}
in view of ${\nabla_{w^{\sf T}}J(w^o)}=0$. Then, (\ref{equ76}) ensures the matrix $\bar{H}=H$ according to the definition in (\ref{equ67}). By referring to (\ref{equ73}), we have
\begin{equation}\label{equ73a}
X=\frac{1}{2}\left(\sum\limits_{k=1}^Nq_k(1-r_k)\right)^{-1}I_M.
\end{equation}

Then, expressions (\ref{equ43}) and (\ref{equ72}) reduce to
\begin{multline}\label{equ43a}
\mathrm{MSD}_{\mathrm{coor},k}= \mathrm{MSD}_{\mathrm{coor,av}}\\=\frac{1}{2}\left(\sum\limits_{k=1}^Nq_k(1-r_k)\right)^{-1}\sum\limits_{k=1}^Nq_k^2\mathrm{Tr}\left(H^{-1}G_k'\right)
\end{multline}
\begin{multline}\label{equ74}
\mathrm{ER}_{\mathrm{coor},k}= \mathrm{ER}_{\mathrm{coor,av}}\\=\frac{1}{4}\left(\sum\limits_{k=1}^Nq_k(1-r_k)\right)^{-1}\sum\limits_{k=1}^Nq_k^2(1-r_k)\mathrm{Tr}\left(G\right).
\end{multline}
We proceed to compare the MSD and ER performance in the stochastic full-gradient and coordinate-descent cases. Let
\begin{align}
\alpha&\triangleq\frac{\sum_{k=1}^Nq_k^2(1-r_k)^2}{\sum_{k=1}^Nq_k(1-r_k)}-\frac{\sum_{k=1}^Nq_k^2}{\sum_{k=1}^Nq_k}\label{equ78}\\
\theta&\triangleq\frac{\sum_{k=1}^Nq_k^2(1-r_k)}{\sum_{k=1}^Nq_k(1-r_k)}-\frac{\sum_{k=1}^Nq_k^2}{\sum_{k=1}^Nq_k}\label{equ79}
\end{align}
and note that $\alpha \leq \theta$, with equality if, and only if, $\{r_k\equiv0\}$.
\begin{corollary}\label{cor3}(\textbf{Performance comparison}).
Under the same conditions of Theorem \ref{theo2}, when the individual costs $J_k(w)$ are identical across the agents, it holds that:

a) if $\alpha\geq0$:
\be\label{app41}
0\leq\mathrm{MSD}_{\mathrm{coor},k}-\mathrm{MSD}_{\mathrm{grad},k}\leq\frac{1}{2}\frac{\theta}{\nu_d}\mathrm{Tr}\left(G\right)
\ee

b) if $\alpha<0$, and $\theta\geq\left(1-{\delta_d}/{\nu_d}\right)\alpha\geq0$:
\begin{multline}\label{equ78a}
0\leq
\mathrm{MSD}_{\mathrm{coor},k}-\mathrm{MSD}_{\mathrm{grad},k}\leq\\\frac{1}{2}\left(\frac{\theta}{\nu_d}+\left(\frac{1}{\delta_d}-\frac{1}{\nu_d}\right)\alpha\right)\mathrm{Tr}\left(G\right)
\end{multline}

c) if $\alpha<0$, and $\theta\leq\left(1-{\nu_d}/{\delta_d}\right)\alpha\leq0$:
\begin{multline}\label{equ84}
\frac{1}{2}\left(\frac{\theta}{\delta_d}+\left(\frac{1}{\nu_d}-\frac{1}{\delta_d}\right)\alpha\right)\mathrm{Tr}\left(G\right)\leq\\
\mathrm{MSD}_{\mathrm{coor},k}-\mathrm{MSD}_{\mathrm{grad},k}\leq0.
\end{multline}

Likewise, it holds that
\be\label{equ80}
\mathrm{ER}_{\mathrm{coor},k}-\mathrm{ER}_{\mathrm{grad},k}=\frac{\theta}{4}\mathrm{Tr}\left(G\right).
\ee
Then, in the case when either the missing probabilities or the quantities $\{q_k\}$ are uniform across the agents, namely, $\{r_k\equiv r\}$ or $\{q_k\equiv q\}$, it follows that
\be\label{equ81}
\mathrm{ER}_{\mathrm{coor},k}=\mathrm{ER}_{\mathrm{grad},k}.
\ee
\end{corollary}

\noindent \emph {Proof}: See Appendix \ref{APP5}.
\hfill $\Box$

Note that for the other choices of parameter $\theta$ that are not indicated in Corollary \ref{cor3}, there is no consistent conclusion on which MSD (between $\mathrm{MSD}_{\mathrm{coor},k}$ and $\mathrm{MSD}_{\mathrm{grad},k}$) is lower.
\begin{figure}
\centering
\includegraphics[width=2.8in]{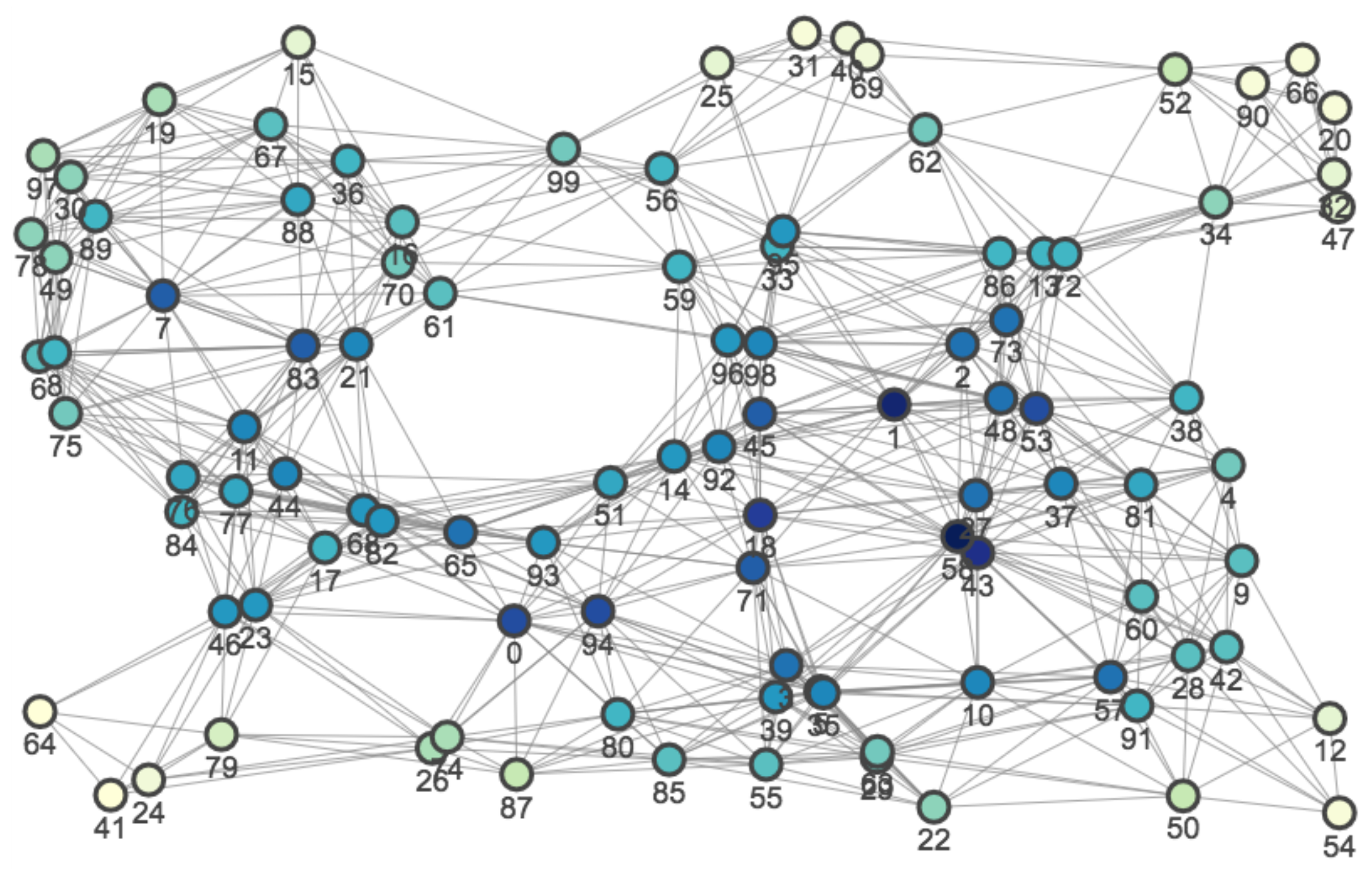}
\caption{Network topology consisting of $N=100$ agents.}
\label{topology100}
\end{figure}
\begin{figure}
	\centering
	\includegraphics[width=2.8in]{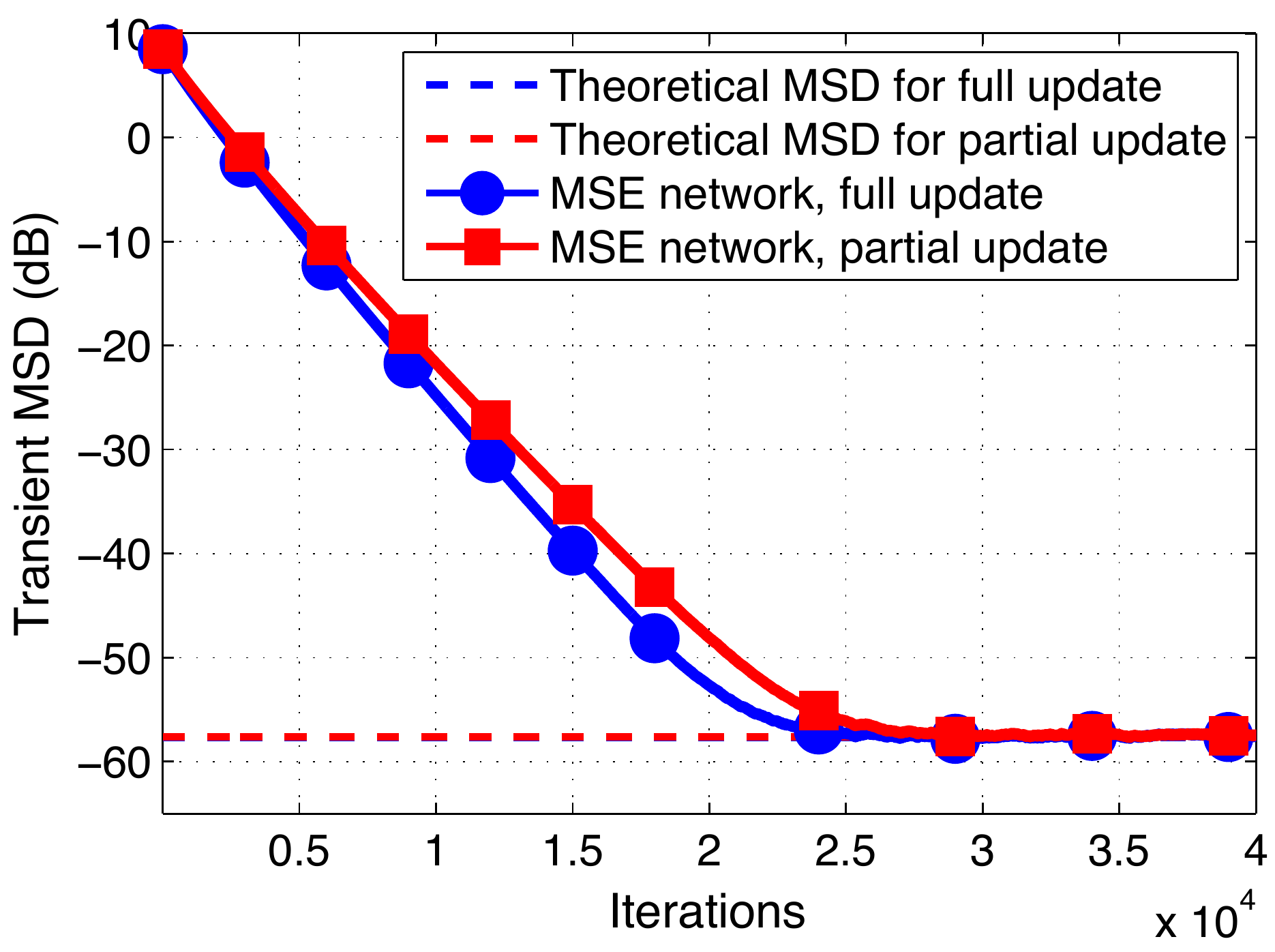}
	\caption{MSD learning curves, averaged over 200 independent runs, in the case of Corollary \ref{cor5} when $\{r_k=0.1\}$. The dashed lines show the theoretical MSD values from (\ref{equ43}).}
	\label{small_probabilities}
\end{figure}
\begin{figure}
	\centering
	\includegraphics[width=2.8in]{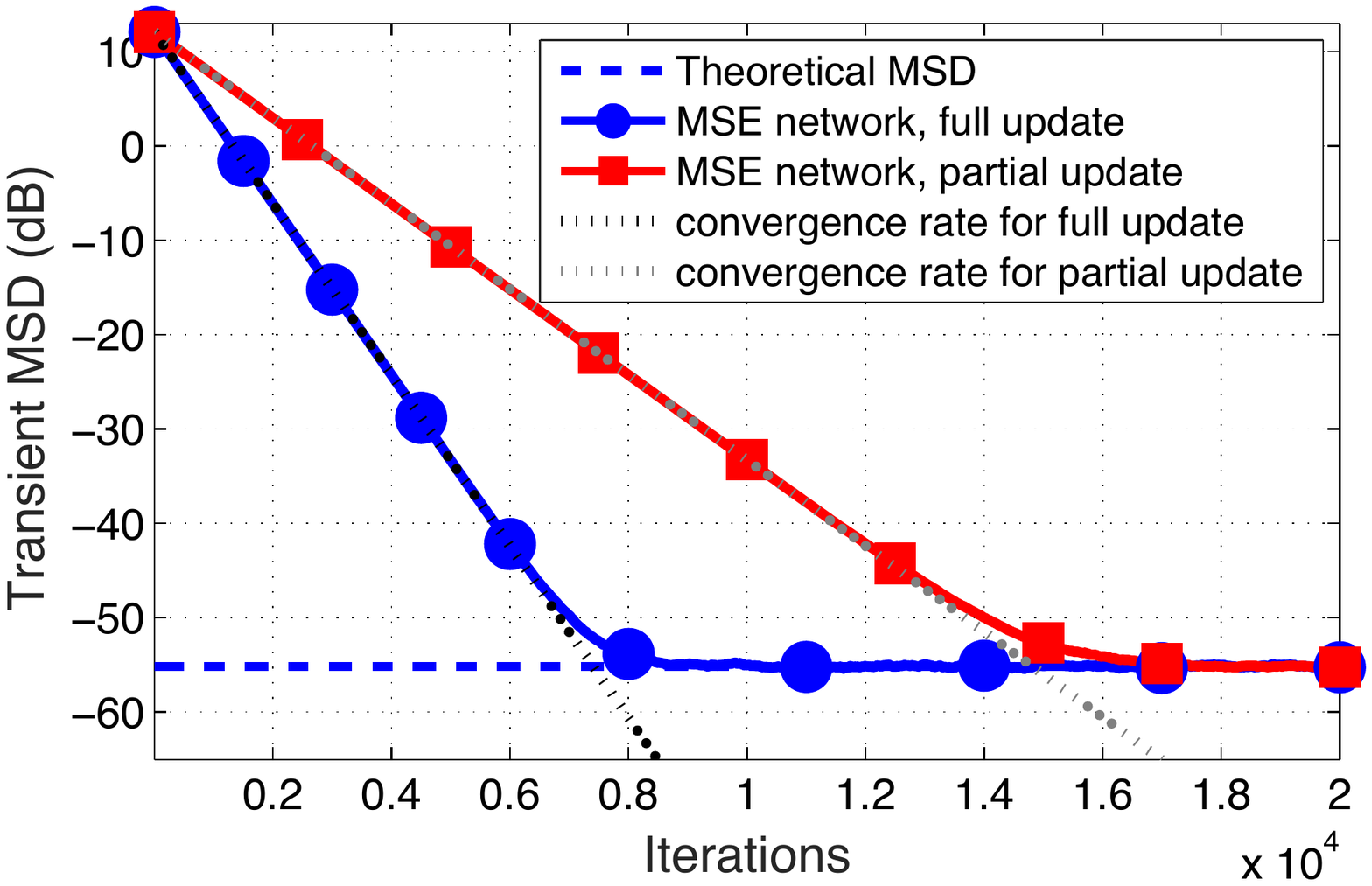}
	\caption{MSD learning curves, averaged over 200 independent runs, in the case of Corollary \ref{cor1} when $\{H_k,G_k\}$ are diagonal. The dashed line along the horizontal axis shows the theoretical MSD value from (\ref{equ43}). Those along the learning curves show the reference recursion at rates formulated by (\ref{equ88}).}
	\label{white_regressors}
\end{figure}
\begin{figure*}
	\centering
	\subfigure[]{\includegraphics[width=2.3in]{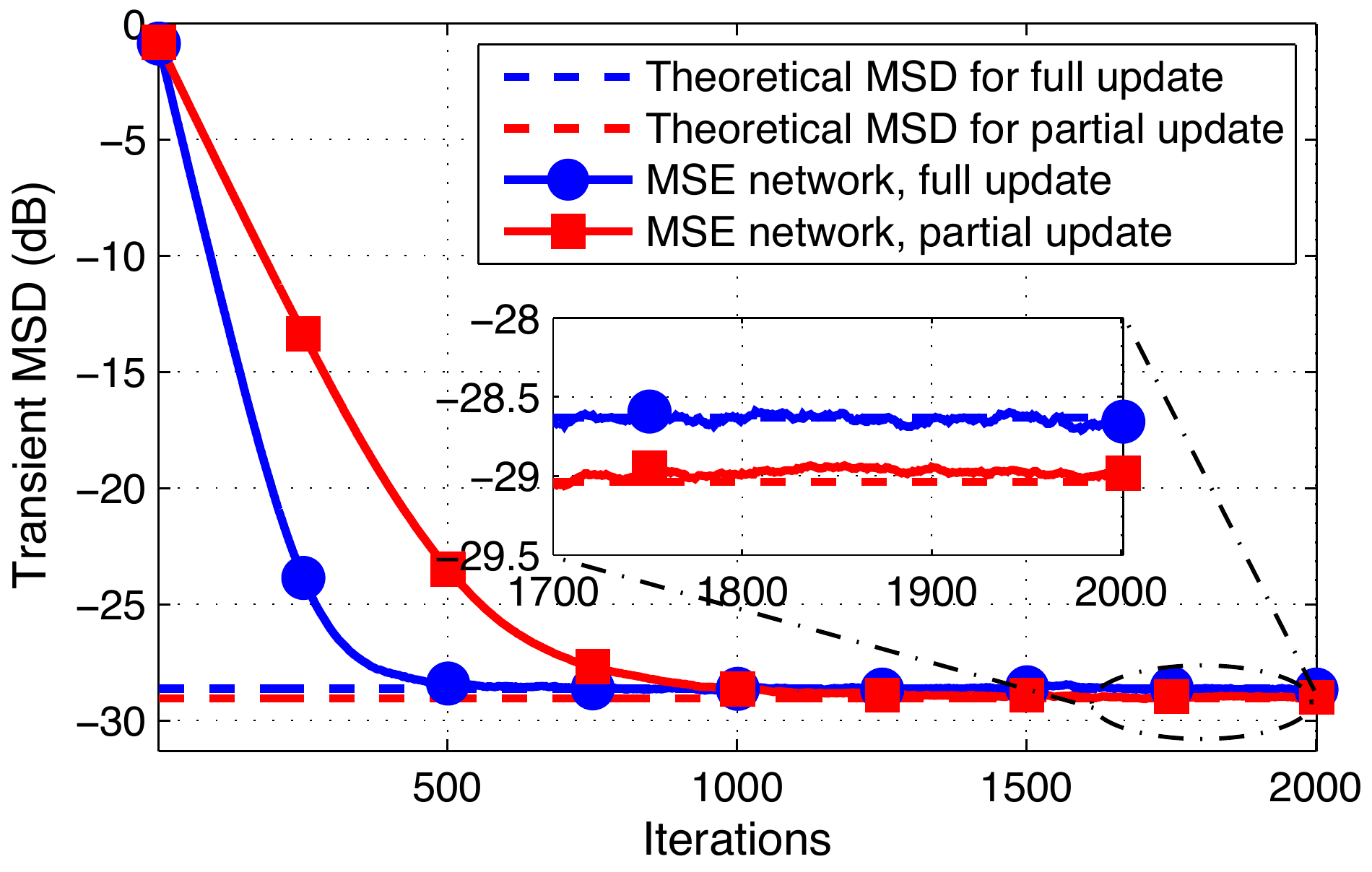}}
	\subfigure[]{\includegraphics[width=2.3in]{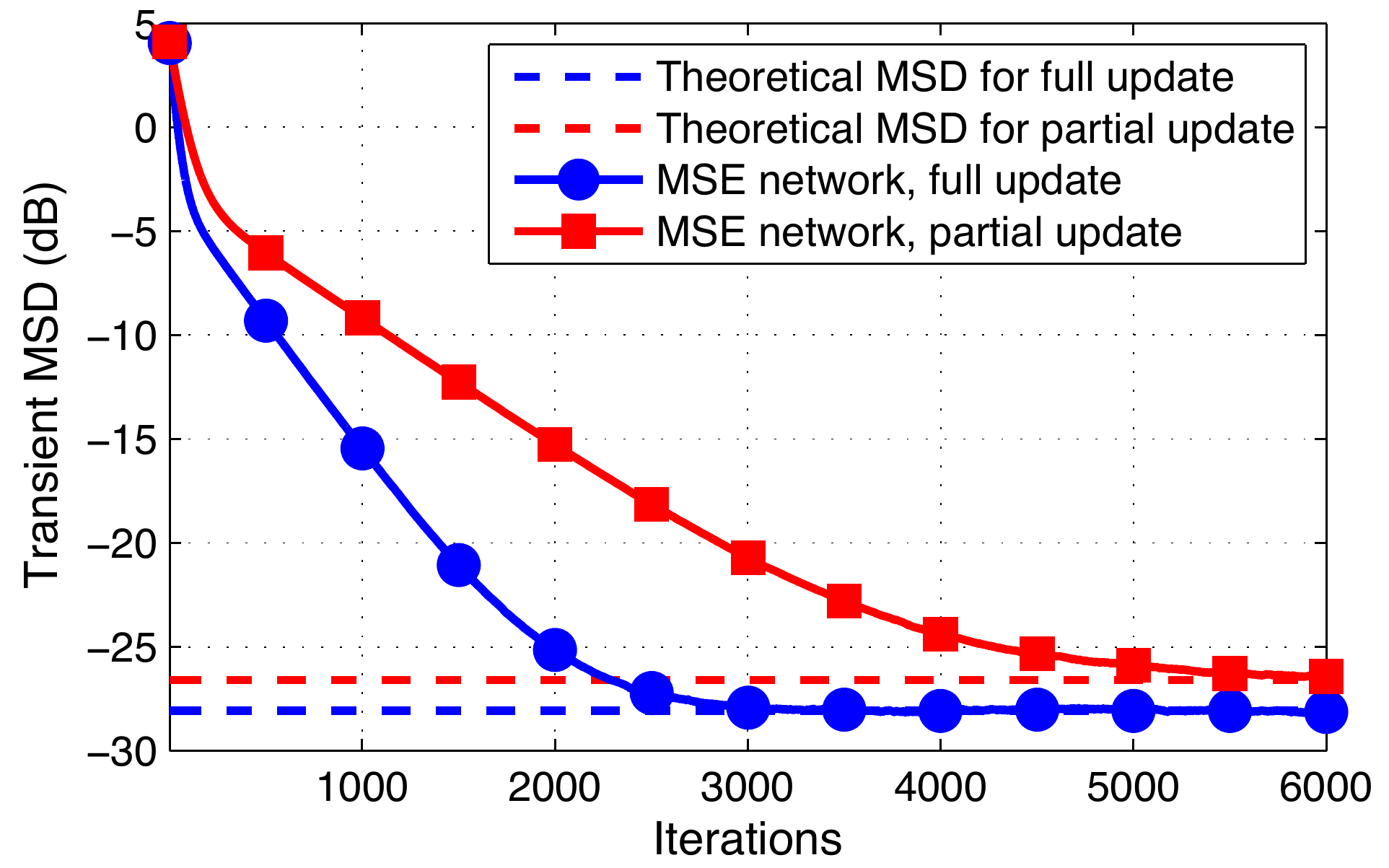}}
	\subfigure[]{\includegraphics[width=2.3in]{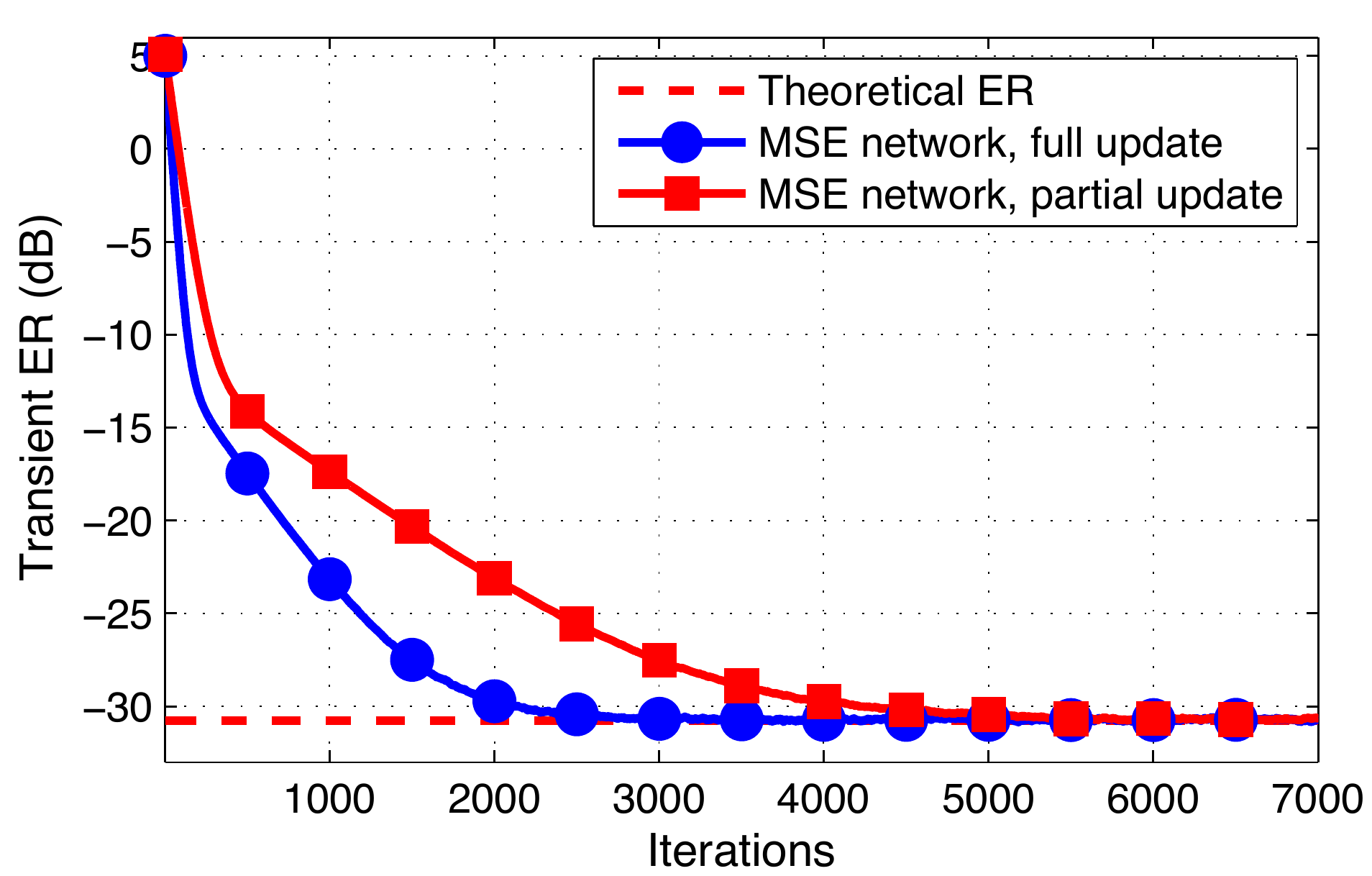}}
	\caption{Learning curves, averaged over 10000 independent runs, and theoretical results calculated from (\ref{equ43}) and (\ref{equ72}) respectively, for a two-agent MSE network, with parameters $\{\pi_1=-0.34,\pi_2=0.99\}$ in (a), and $\{\pi_1=0.34,\pi_2=0.99\}$ in (b) and (c).}
	\label{two_agents}
\end{figure*}
\section{Simulation Results}
In this section, we illustrate the results by considering MSE networks and logistic regression networks; both settings satisfy condition (\ref{equ6}) and  Assumptions \ref{ass2} through \ref{ass5}.
\subsection{MSE Networks}
In the following examples, we will test performance of the associated algorithms in the case when uniform missing probabilities are utilized across the agents. We adopt the ATC
formulation, and set the combination matrices $A_1=I$, and $A_2$ according to the averaging rule \cite[p. 664]{ASayed2014} in the first two examples, and Metropolis rule \cite[p. 664]{ASayed2014} in the third example.
In the first example, we test the case when the gradient vectors are missing with small probabilities across the agents. Figure \ref{topology100} shows a network topology with $N = 100$ agents. The parameter vector $w^o$ is randomly generated with $M=10$. The regressors are generated by the first-order autoregressive model
\begin{equation}\label{equ66}
\bm u_{k,i}(m) = \pi_k\bm u_{k,i}(m-1)+\sqrt{1-\pi_k^2}\bm t_{k,i}(m),\,1\leq m<M
\end{equation}
and the variances are scaled to be 1. The processes $\{\bm t_{k,i}\}$ are zero-mean, unit-variance, and independent and identically distributed (\emph{i.i.d}) Gaussian sequences. The parameters $\{\pi_k\}$ are generated from a uniform distribution on the interval $(-1,1)$. The noises, uncorrelated with the regression vectors, are zero-mean white Gaussian sequences with the variances uniformly distributed over $(0.001,0.1)$. The step-sizes $\{\mu_k\}$ across the agents are generated from a uniform distribution on the interval $(0.0001,0.0005)$. We choose a small missing probability $\{r_k=0.1\}$. Figure \ref{small_probabilities} shows the simulation results, which are averaged over 200 independent runs, as well as the theoretical MSD values calculated from (\ref{equ43}), which are $-57.72$dB and $-57.61$dB, respectively, for the full and partial update case. It is clear from the figure that, when the gradient information is missing with small probabilities, the performance of the coordinate-descent case is close to that of the full-gradient diffusion case.

In the second example, we test the case when the regressors are white across the agents. We randomly generate $w^o$ of size $M=10$. The white regressors are generated from zero-mean white Gaussian sequences, and the powers, which vary from entry to entry, and from agent to agent, are uniformly distributed over $(0.05,0.15)$. The noises $\{\bm v_k(i)\}$, uncorrelated with the regressors, are zero-mean white Gaussian sequences, with the variances $\{\sigma_{v,k}^2\}$ generated from uniform distribution on the interval $(0.0001,0.01)$. The step-sizes are uniformly distributed over $(0.001,0.01)$. The results, including the theoretical MSD value from (\ref{equ43}) in Theorem \ref{theo2}, the simulated MSD learning curves, and the convergence rates from (\ref{equ88}), are illustrated by Fig. \ref{white_regressors}, where the results are averaged over 200 independent runs. It is clear from the figure that, when white regressors are utilized in MSE networks, the stochastic coordinate-descent case converges to the same MSD level as the full-gradient diffusion case, which verifies (\ref{equ52h}), at a convergence rate formulated in (\ref{equ88}).

In the third example, we revisit the two-agent MSE network discussed in Appendix \ref{APP6}, i.e., $N=2$.
We randomly generate $w^o$ of size $M=2$. The step-sizes $\mu_1=\mu_2=0.005$ are uniform across the agents, which gives $q_1=q_2=2.5\times10^{-3}$. The missing probabilities $r_1=r_2=0.5$. The noises $\{\bm v_1(i),\bm v_2(i)\}$ are zero-mean white Gaussian sequences with the variances $\{\sigma_{v,1}^2=0.5,\sigma_{v,2}^2=5\times10^{-4}\}$. The regressors, uncorrelated with the noise sequences, are scaled such that the covariance matrices are of the form
\be\label{equ54}
R_{u,1}=\left[\begin{array}{cc}|\pi_1|&\pi_1\\ \pi_1&1\end{array}\right],\,R_{u,2}=\left[\begin{array}{cc}|\pi_2|&\pi_2\\ \pi_2&1\end{array}\right]
\ee
with $|\pi_1|<1, |\pi_2|<1$. Now we select parameters $\{\pi_1=-0.34,\pi_2=0.99\}$, which satisfy condition (\ref{equ86}), and $\{\pi_1=0.34,\pi_2=0.99\}$ to illustrate the cases of $\mathrm{MSD}_{\mathrm{coor},k}<\mathrm{MSD}_{\mathrm{grad},k}$ and $\mathrm{MSD}_{\mathrm{coor},k}>\mathrm{MSD}_{\mathrm{grad},k}$ respectively. Fig. \ref{two_agents} (a) shows the simulation results with the parameters $\{\pi_1=-0.34,\pi_2=0.99\}$. Figures \ref{two_agents} (b) and \ref{two_agents} (c) show the simulation results with the parameters $\{\pi_1=0.34,\pi_2=0.99\}$. All results are averaged over 10000 independent runs.
It is clear from the figures that the simulation results match well with the theoretical results from Theorem \ref{theo2}. In Fig. \ref{two_agents} (a), the steady-state MSD of the stochastic coordinate-descent case is slightly lower than that of the full-gradient diffusion case, by about $0.32$dB, which is close to the theoretical MSD difference of $0.41$dB from (\ref{equ64}). The MSD performance is better in the full-gradient diffusion case in Fig. \ref{two_agents} (b), and the difference between these two MSDs at steady state is $1.71$dB, which is close to the theoretical difference of $1.49$dB from (\ref{equ64}). The ER performance for both the stochastic coordinate-descent and full-gradient diffusion cases are the same as illustrated in Fig. \ref{two_agents} (c), which verifies the theoretical result in (\ref{equ730b}).


\begin{figure}
	\centering
	\includegraphics[width=2.4in]{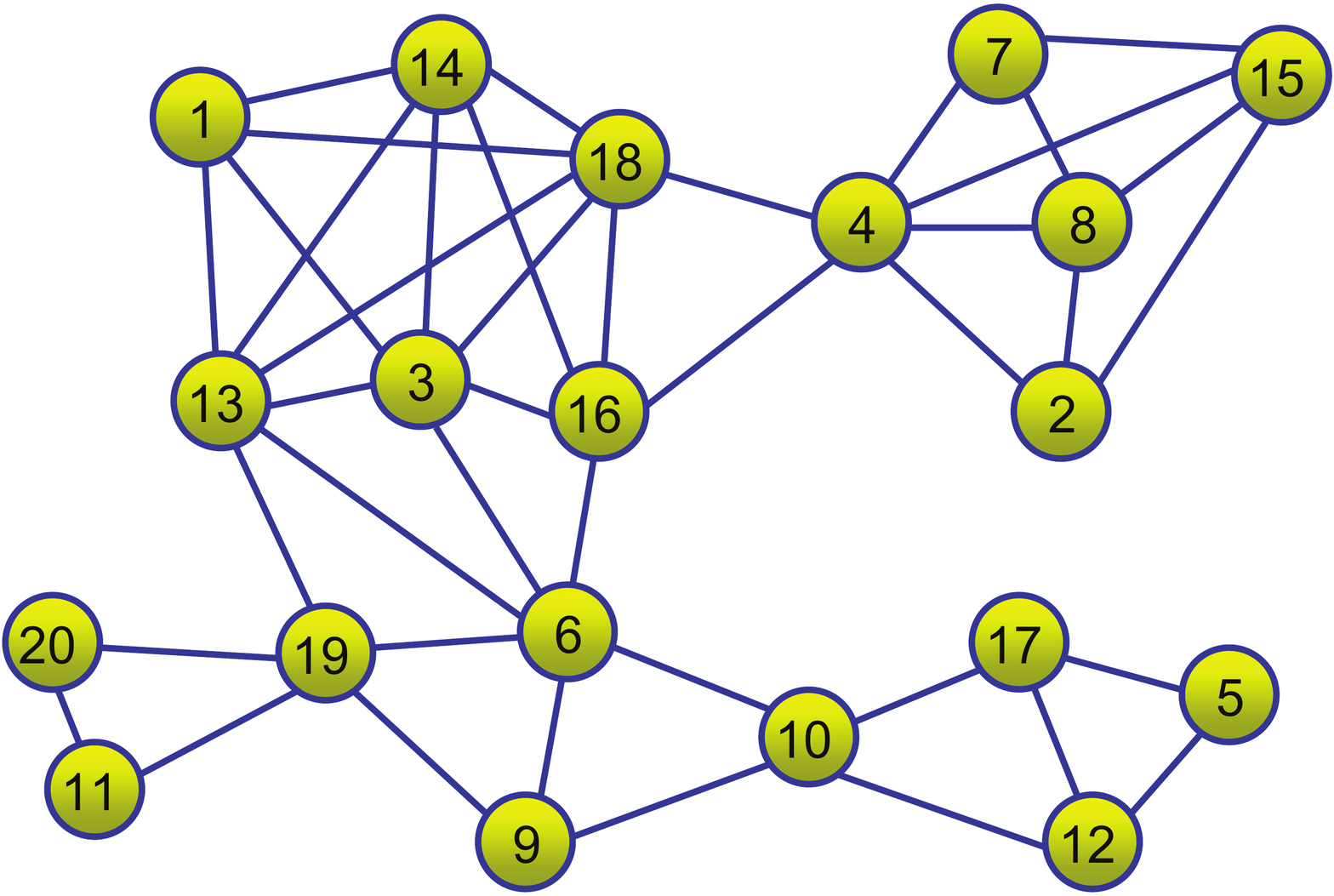}
	\caption{Network topology consisting of $N=20$ agents.}
	\label{topology}
\end{figure}
\begin{figure*}
	\centering
	\subfigure[]{\includegraphics[width=2.3in]{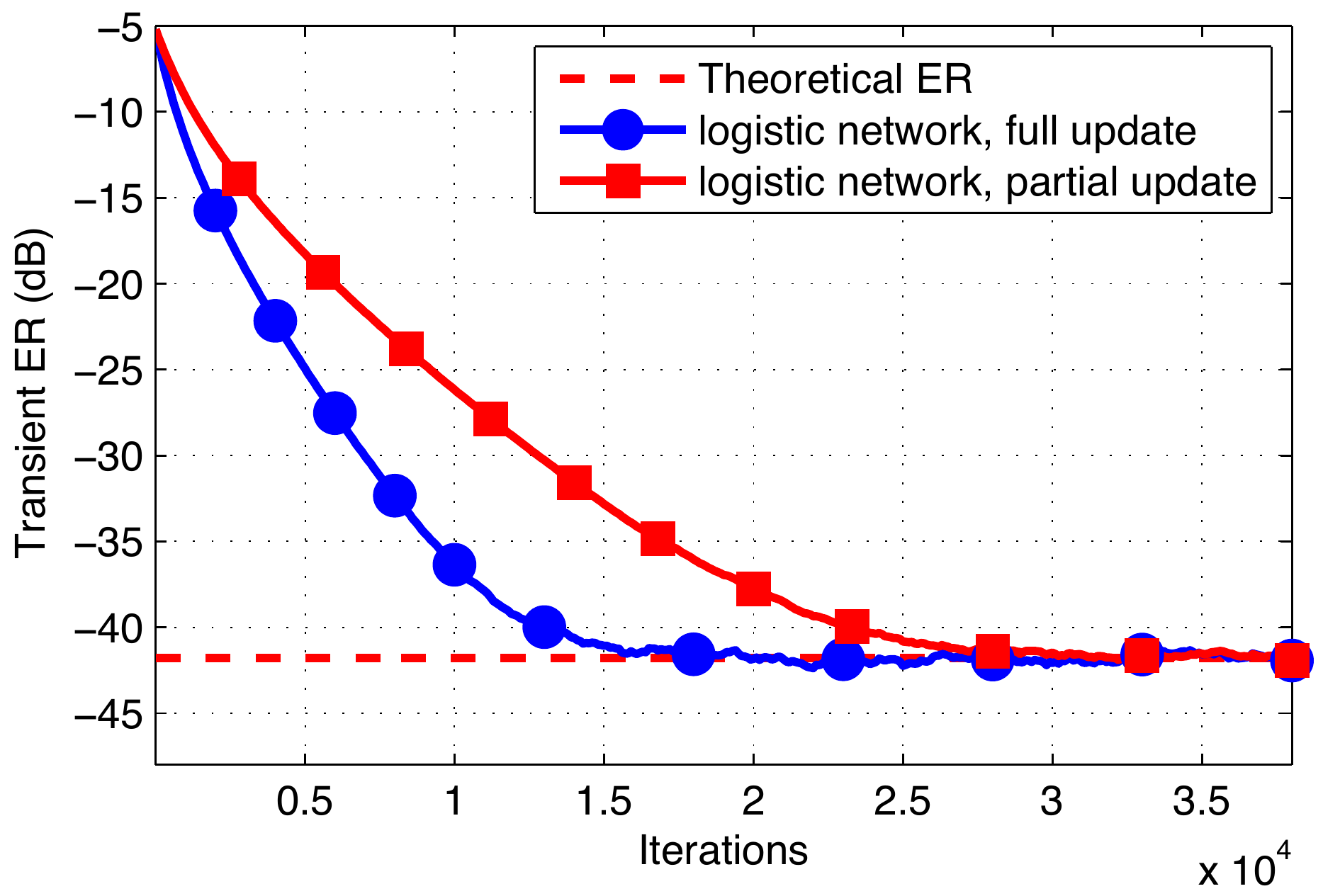}}
	\subfigure[]{\includegraphics[width=2.3in]{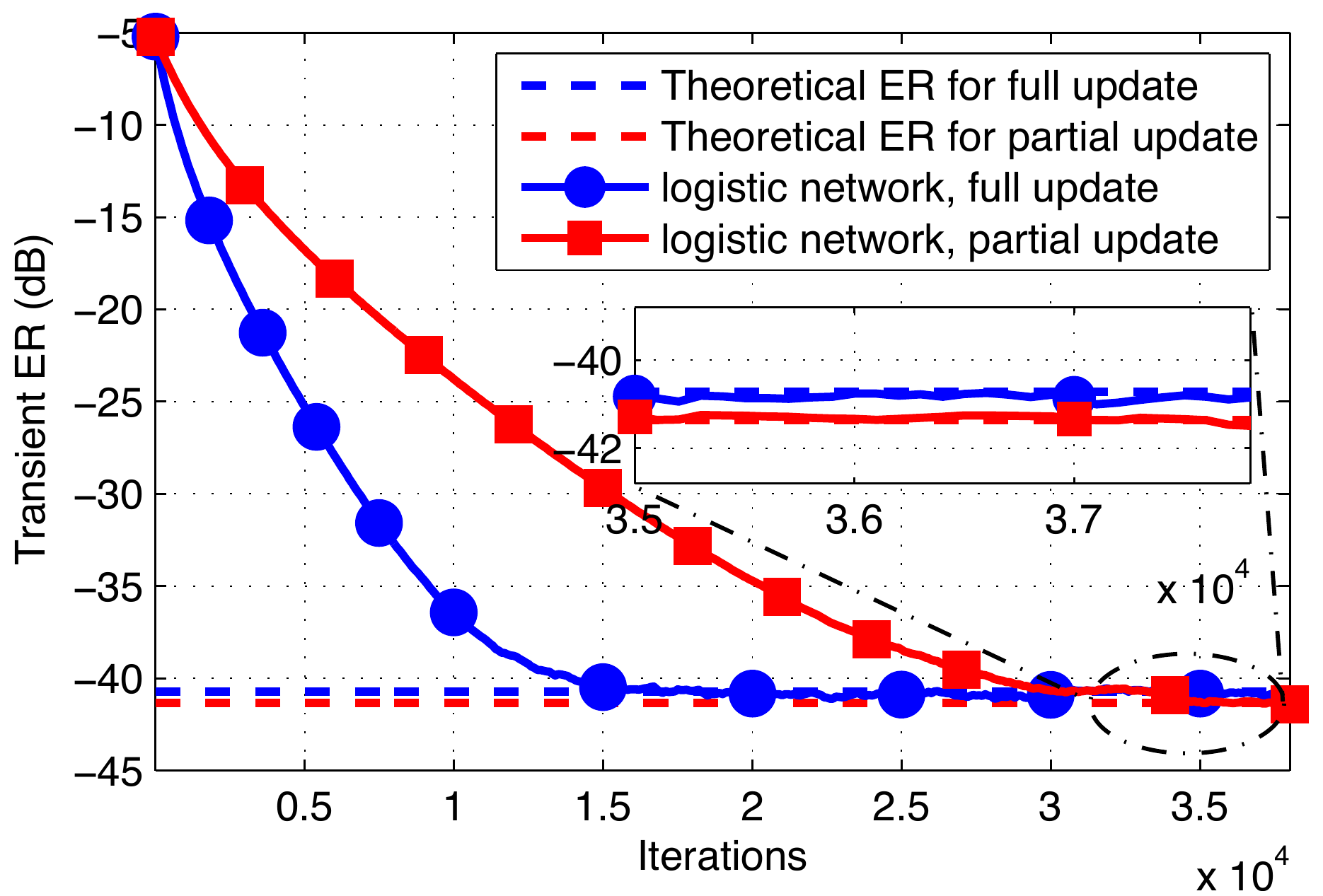}}
	\subfigure[]{\includegraphics[width=2.3in]{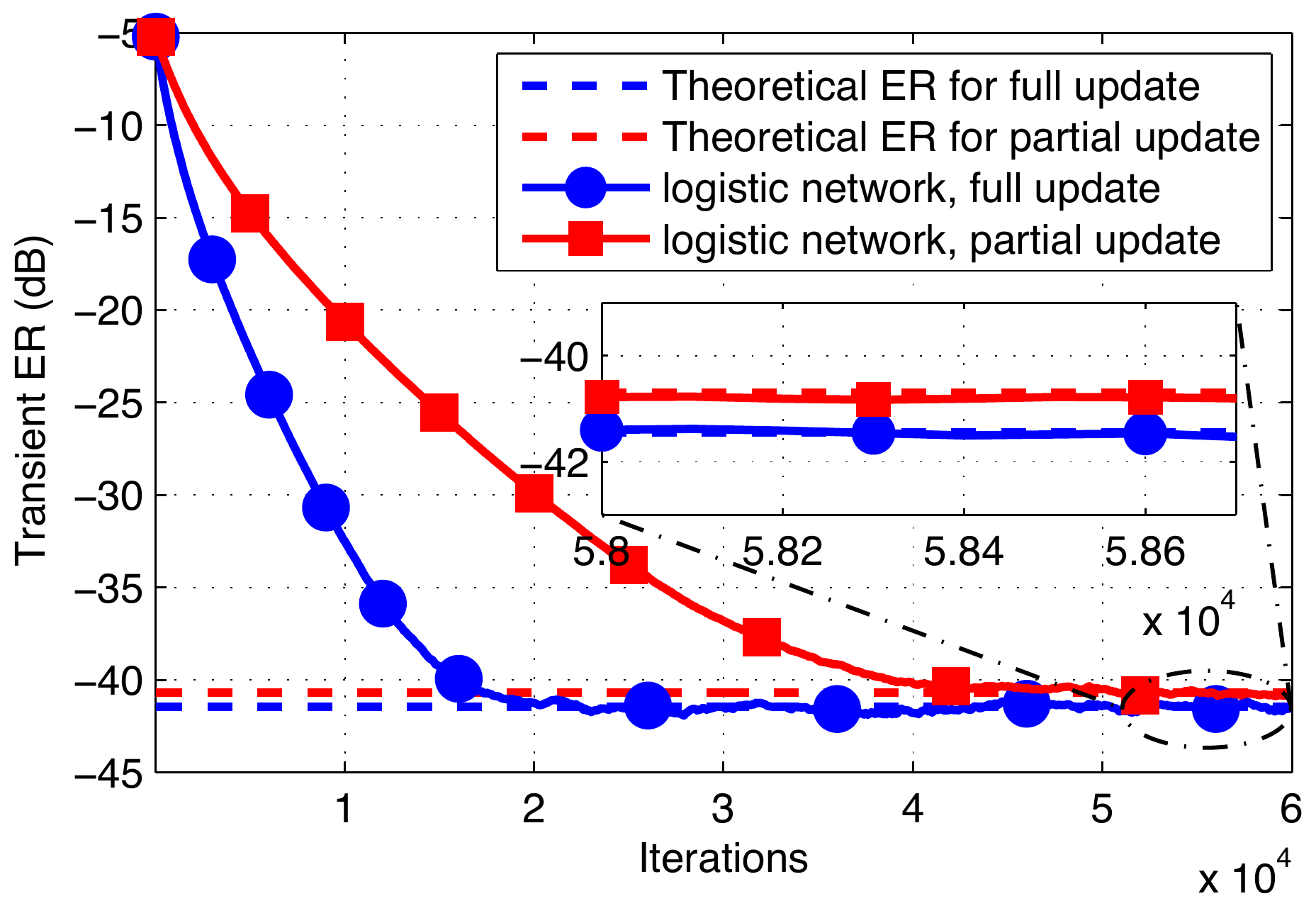}}
	\caption{ER learning curves, averaged over 1000 independent runs, and theoretical results from (\ref{equ74}) for diffusion learning over a logistic network with full or partial updates. Corollary \ref{cor3} is tested in (a) when a uniform step-size and a doubly-stochastic combination matrix are utilized across the network. Corollary \ref{cor3} is tested when the parameters $\{\mu_k\}$ and $\{r_k\}$ are scaled to make $\theta$ in (\ref{equ79}) negative in (b) and positive in (c).}
	\label{logistic}
\end{figure*}
\subsection{Logistic Networks}
We now consider an application in the context of pattern classification. We assign with each agent the logistic risk
\begin{equation}\label{equ50}
J_k(w)=\frac{\rho}{2}\|w\|^2 + \mathbb{E}\left\{\ln\left[1+e^{-\bm \gamma_k(i)\bm h^{\sf T}_{k,i}w}\right]\right\}
\end{equation}
with regularization parameter $\rho>0$, and where the labels $\{\bm \gamma_k(i)=\pm 1\}$ are binary random and the $\{\bm h_{k,i}\}$ are feature vectors. The objective is for the agents to determine a parameter vector $w^o$ to enable classification by estimating the class labels via $\widehat{\bm \gamma}_{k}(i)=\bm h_{k,i}^{\sf T} w^o$.

We proceed to test the theoretical findings in Corollary \ref{cor3}. Consider the network topology in Fig. \ref{topology} with $N=20$ agents. We still adopt the ATC
formulation, and set the combination matrices $A_1=I$, and $A_2$ according to the Metropolis rule in \cite[p. 664]{ASayed2014}. The feature vectors and the unknown parameter vector are randomly generated from uncorrelated zero-mean unit-variance \emph{i.i.d} Gaussian sequences, both of size $M = 10$. The parameter $\rho$ in (\ref{equ50}) is set to 0.01. To generate the trajectories for the experiments, the optimal solution to (\ref{equ50}), $w^o$, the Hessian matrix $H$, and the gradient noise covariance matrix, $G$, are first estimated off-line by applying a batch algorithm to all data points.

In the first example, we consider the case when a uniform step-size $\{\mu_k = 0.005\}$ is utilized across the agents. All entries of the stochastic gradient vectors are missing completely at random with positive probabilities that are uniformly distributed over $(0,1)$. Figure \ref{logistic} (a) shows the transient ER curves for the diffusion strategies with complete and partial gradients, where the results are averaged over 1000 independent runs. The figure also shows the theoretical result calculated from (\ref{equ74}). It is clear from Figure \ref{logistic} (a) that the same ER performance is obtained in the stochastic coordinate-descent and full-gradient diffusion cases, by utilizing a uniform step-size and a doubly-stochastic combination matrix across the agents (in which case the parameters $\{q_k\}$ in (\ref{equ11}) are identical across the agents), which is in agreement with the theoretical analysis in (\ref{equ81}).

In the second and third examples, we randomly generate the step-sizes $\{\mu_k\}$ and missing probabilities $\{r_k\}$ by following uniform distributions on the intervals $(0.001,0.01)$ and $(0,1)$ respectively. In Figure \ref{logistic} (b), the parameters $\{\mu_k\}$ and $\{r_k\}$ are scaled to get a negative value for $\theta$ in (\ref{equ79}), and in Fig. \ref{logistic} (c), those parameters are scaled to make $\theta$ positive. Figures \ref{logistic} (b) and \ref{logistic} (c) show respectively the transient ER learning curves in these two cases for the diffusion strategies with complete and partial gradients, where the results are averaged over 1000 independent runs. The figures also show the theoretical results calculated from (\ref{equ74}). It is clear from Figs. \ref{logistic} (b) and \ref{logistic} (c) that these learning curves converge to their theoretical results at steady state. In Fig. \ref{logistic} (b) where $\theta<0$, the stochastic coordinate-descent case converges to a lower ER level than the full-gradient diffusion case, and the difference between these two ERs is $0.637$dB, which is close to the theoretical difference of $0.640$dB from (\ref{equ80}). In Fig. \ref{logistic} (c) where $\theta>0$, the steady-state ER in the full-gradient diffusion case is lower than that of the stochastic coordinate-descent case, by about $0.726$dB, which is close to the theoretical difference of $0.750$dB from (\ref{equ80}).

\appendices
\section{Proof of Theorem 1}\label{APP1}
Let $P=A_1A_2$. It was argued in \cite[p.510]{ASayed2014} that $P$ admits a Jordan canonical decomposition of the form
$P=V_\epsilon J V_\epsilon^{-1}$
where
\be
V_\epsilon\triangleq\left[\begin{array}{cc}p&V_R\end{array}\right],\,
V_\epsilon^{-1}\triangleq\left[\begin{array}{c}\mathds{1}^{\sf{T}}\\V_L^{\sf{T}}\end{array}\right],\,
J=\left[\begin{array}{cc}1&0\\0&J_\epsilon\end{array}\right]\label{app4}
\ee
$p$ is defined by (\ref{equ11a}), $\epsilon$ denotes an arbitrary positive scalar that we are free to choose, and the matrix $J_{\epsilon}$ has a Jordan structure with $\epsilon$ appearing in the first lower diagonal rather than unit entries.
All eigenvalues of $J_{\epsilon}$ are strictly inside the unit circle. Then,
\begin{align}\label{app6}
\mathcal{P}\triangleq P\otimes I_{M}
\triangleq\mathcal{V}_\epsilon\mathcal{J}\mathcal{V}_\epsilon^{-1}
\end{align}
where
$\mathcal{V}_\epsilon\triangleq{V}_\epsilon\otimes I_{M},\,\mathcal{J}\triangleq J\otimes I_{M}$.
Using (\ref{app6}), we can rewrite $\bm{\mathcal{B}}_{i}$ from (\ref{equ25}) as
\begin{align}\label{app7}
\bm{\mathcal{B}}_{i}
\triangleq\left(\mathcal{V}_\epsilon^{-1}\right)^{{\sf{T}}}(\mathcal{J}^{{\sf{T}}}-\bm{\mathcal{D}}_{i}^{{\sf{T}}})\mathcal{V}_\epsilon^{\sf{T}}
\end{align}
where
\begin{align}\label{app8}
\bm{\mathcal{D}}_{i}^{\sf{T}}&\triangleq\mathcal{V}_\epsilon^{{\sf{T}}}\mathcal{A}_2^{{\sf{T}}}\mathcal{M}\bm{\Gamma}_i\bm{\mathcal{H}}_{i-1}\mathcal{A}_1^{{\sf{T}}}\left(\mathcal{V}_\epsilon^{-1}\right)^{\sf{T}}\nonumber\\
&=\left[\begin{array}{cc}\bm{D}_{11,i}^{\sf{T}}&\bm{D}_{21,i}^{\sf{T}}\\\bm{D}_{12,i}^{\sf{T}}&\bm{D}_{22,i}^{\sf{T}}\end{array}\right]
\end{align}
and
\be
\bm{D}_{11,i}=\sum\limits_{k=1}^Nq_k\bm{H}_{k,i-1}\bm{\Gamma}_{k,i}\label{app9}
\ee
with the vector $q=\{q_k\}$ defined by (\ref{equ11}). With regards to the norm of $\bm{D}_{11,i}$, we observe that contrary to the arguments in \cite[p. 511]{ASayed2014}, this matrix is not symmetric anymore in the coordinate-descent case due to the presence of $\bm{\Gamma}_{k,i}$. We therefore need to adjust the arguments, which we do next.

Let
\begin{align}\label{app35}
\bm {\bar{D}}_{11,i}&\triangleq\mathbb{E}\left[\bm D_{11,i}|\bm{\mathcal{F}}_{i-1}\right]\nn
&=\sum_{k=1}^Nq_k\bm{H}_{k,i-1}\mathbb{E}\left[\bm\Gamma_{k,i}\right]\nn
&\stackrel{(\ref{equ4a})}=\sum_{k=1}^Nq_k(1-r_k)\bm{H}_{k,i-1}\nn
&=\mathbb{E}\left[\bm D^{\sf{T}}_{11,i}|\bm{\mathcal{F}}_{i-1}\right].
\end{align}
Noting that
\be\label{app36}
\mathbb{E}\left[\bm\Gamma_{k,i}\bm\Gamma_{j,i}\right]=\left\{\begin{array}{cl}(1-r_k)(1-r_j),&k\neq j\\1-r_k,&k=j\end{array}\right.
\ee
we introduce
\begin{align}
\bm R_{D_{11},i}&\triangleq\mathbb{E}\left[\left(\bm D_{11,i}-\bm {\bar{D}}_{11,i}\right)\left(\bm D_{11,i}-\bm {\bar{D}}_{11,i}\right)^{\sf{T}}|\bm{\mathcal{F}}_{i-1}\right]\nn
&{\hspace{0.1cm}=\mathbb{E}\left[\bm D_{11,i}\bm D^{\sf{T}}_{11,i}|\bm{\mathcal{F}}_{i-1}\right]-\bm {\bar{D}}_{11,i}\mathbb{E}\left[\bm D^{\sf{T}}_{11,i}|\bm{\mathcal{F}}_{i-1}\right]-}\nn
&{\hspace{0.5cm}\mathbb{E}\left[\bm D_{11,i}|\bm{\mathcal{F}}_{i-1}\right]\bm {\bar{D}}_{11,i}+\bm {\bar{D}}^2_{11,i}}\nn
&\hspace{-0.17cm}\stackrel{(\ref{app35})}=\mathbb{E}\left[\bm D_{11,i}\bm D^{\sf{T}}_{11,i}|\bm{\mathcal{F}}_{i-1}\right]-\bm {\bar{D}}^2_{11,i}\label{app38}\\
&=\sum_{k=1}^N\sum_{j=1}^Nq_kq_j\bm {H}_{k,i-1}\mathbb{E}\left[\bm \Gamma_{k,i}\bm \Gamma_{j,i}\right]\bm {H}_{j,i-1}-\nn
&\hspace{0.5cm}\sum_{k=1}^N\sum_{j=1}^Nq_kq_j(1-r_k)(1-r_j)\bm {H}_{k,i-1}\bm {H}_{j,i-1}\nn
&\hspace{-0.15cm}\stackrel{(\ref{app36})}=\sum_{k=1}^N\sum_{j\neq k=1}^Nq_kq_j(1-r_k)(1-r_j)\bm {H}_{k,i-1}\bm {H}_{j,i-1}-\nn
&\hspace{0.5cm}\sum_{k=1}^N\sum_{j=1}^Nq_kq_j(1-r_k)(1-r_j)\bm {H}_{k,i-1}\bm {H}_{j,i-1}
+\nonumber\\
&\hspace{0.5cm}\sum_{k=1}^Nq_k^2(1-r_k)\bm {H}_{k,i-1}^2\nn
&=\sum_{k=1}^Nq_k^2(1-r_k)\bm {H}_{k,i-1}^2-\sum_{k=1}^Nq_k^2(1-r_k)^2\bm {H}_{k,i-1}^2\nn
&=\sum_{k=1}^Nq_k^2(1-r_k)r_k\bm {H}_{k,i-1}^2.\label{app37}
\end{align}
Recall, from (\ref{equ6}) and (\ref{equ27}), that
\be\label{app53}
0<\nu_d I_M\leq\bm {H}_{k,i-1}\leq\delta_d I_M.
\ee
Then, matrices $\bm {\bar{D}}_{11,i}$ and $\bm R_{D_{11},i}$ are symmetric positive-definite. Following similar arguments to those in \cite[pp. 511--512]{ASayed2014}, we have
\be\label{app39}
\|I_M-\bm {\bar{D}}_{11,i}\|\leq1-\sigma_{11}\mu_{\max},\,\|\bm R_{D_{11},i}\|\leq\beta_{11}^2\mu_{\max}^2
\ee
for some positive constants $\sigma_{11}$ and $\beta^2_{11}$, and sufficiently small $\mu_{\max}$.

Now, multiplying both sides of (\ref{equ24}) by $\mathcal{V}_\epsilon^{\sf{T}}$, we have
\begin{equation}\label{app19}
\mathcal{V}^{\sf{T}}_\epsilon\widetilde{\bm{w}}_i = (\mathcal{J}^{{\sf{T}}}-\bm{\mathcal{D}}_{i}^{{\sf{T}}})\mathcal{V}^{\sf{T}}_\epsilon\widetilde{\bm{w}}_{i-1}+\mathcal{V}^{\sf{T}}_\epsilon\mathcal{A}_2^{{\sf{T}}}\mathcal{M}\bm{\Gamma}_i\bm{s}_i
\end{equation}
where (\ref{app7}) was used. Let
\begin{equation}
\mathcal{V}^{\sf{T}}_\epsilon{\widetilde{{\bm{w}}}}_i=\left[\begin{array}{c}\left(p^{\sf{T}}\otimes I_{M}\right)\widetilde{\bm{w}}_i\\\left(V_R^{\sf{T}}\otimes I_{M}\right)\widetilde{\bm{w}}_i\end{array}\right]\triangleq\left[\begin{array}{c}\bar{{\bm{w}}}_i\\\check{{\bm{w}}}_i\end{array}\right]\label{app19a}
\end{equation}
\begin{equation}
\mathcal{V}^{\sf{T}}_\epsilon\mathcal{A}_2^{{\sf{T}}}\mathcal{M}\bm{\Gamma}_i\bm{s}_i=\left[\begin{array}{c}\left(q^{\sf{T}}\otimes I_{M}\right)\bm{\Gamma}_i\bm{s}_i\\\left(V_R^{\sf{T}}\otimes I_{M}\right)\mathcal{A}_2^{{\sf{T}}}\mathcal{M}\bm{\Gamma}_i\bm{s}_i\end{array}\right]
\triangleq\left[\begin{array}{c}\bar{{\bm{s}}}_i\\\check{{\bm{s}}}_i\end{array}\right]\label{app19b}
\end{equation}
We then rewrite (\ref{app19}) as
\begin{align}\label{app20}
\hspace{-0.3cm}
\left[\begin{array}{c}\bar{{\bm{w}}}_i\\\check{{\bm{w}}}_i\end{array}\right]=\left[\begin{array}{cc}I_{M}-\bm D^{\sf{T}}_{11,i}&-\bm D^{\sf{T}}_{21,i}\\-\bm D_{12,i}^{\sf{T}}&\mathcal{J}^{\sf{T}}_\epsilon-\bm D_{22,i}^{\sf{T}}\end{array}\right]\left[\begin{array}{c}\bar{{\bm{w}}}_{i-1}\\\check{{\bm{w}}}_{i-1}\end{array}\right]+\left[\begin{array}{c}\bar{{\bm{s}}}_i\\\check{{\bm{s}}}_i\end{array}\right]
\end{align}
where the asymmetry of the matrix $\bm{D}_{11,i}$ in this case leads to a difference in the first row, compared to the arguments in \cite[pp. 514--515]{ASayed2014}. We adjust the arguments as follows. Using Jensen's inequality, we have \cite[p. 515]{ASayed2014}:
\begin{multline}\label{app021}
\mathbb{E}[\|\bar{{\bm{w}}}_i\|^2|\bm{\mathcal{F}}_{i-1}]\leq\frac{1}{1-t}\mathbb{E}[\|(I_{M}-\bm D^{\sf{T}}_{11,i})\bar{{\bm{w}}}_{i-1}\|^2|\bm{\mathcal{F}}_{i-1}]\\+\frac{1}{t}\mathbb{E}[\|\bm D^{\sf{T}}_{21,i}\check{{\bm{w}}}_{i-1}\|^2|\bm{\mathcal{F}}_{i-1}]+\mathbb{E}[\|\bar{\bm s}_i\|^2|\bm{\mathcal{F}}_{i-1}]
\end{multline}
for any $0<t<1$, where the expectation of the cross term between $\bar{\bm s}_i$ and $(I_{M}-\bm D^{\sf{T}}_{11,i})\bar{{\bm{w}}}_{i-1}-\bm D^{\sf{T}}_{21,i}\check{{\bm{w}}}_{i-1}$ vanishes conditioned on $\bm{\mathcal{F}}_{i-1}$ and $\bm{\Gamma}_i$ in view of (\ref{equ20}), and the result in (\ref{app021}) follows by  taking the expectations again on both sides over $\bm{\Gamma}_i$. Then, the first term on the right hand side of (\ref{app021}) can be bounded by
\begin{align}
&\mathbb{E}\left[\|(I_{M}-\bm D^{\sf{T}}_{11,i})\bar{{\bm{w}}}_{i-1}\|^2|\bm{\mathcal{F}}_{i-1}\right]\nonumber\\
&\hspace{0.1cm}=\left(\bar{{\bm{w}}}_{i-1}\right)^{\sf{T}}\mathbb{E}\left[\left(I_{M}-\bm D_{11,i}\right)\left(I_{M}-\bm D^{\sf{T}}_{11,i}\right)|\bm{\mathcal{F}}_{i-1}\right]\bar{{\bm{w}}}_{i-1}\nonumber\\
&\hspace{0.05cm}\stackrel{(a)}\leq\lambda_{\max}\left(\mathbb{E}\left[\left(I_{M}-\bm D_{11,i}\right)\left(I_{M}-\bm D^{\sf{T}}_{11,i}\right)|\bm{\mathcal{F}}_{i-1}\right]\right)\left\|\bar{{\bm{w}}}_{i-1}\right\|^2\nn
&\hspace{0.05cm}\stackrel{(b)}=\big\|\mathbb{E}\left[\left(I_{M}-\bm D_{11,i}\right)\left(I_{M}-\bm D^{\sf{T}}_{11,i}\right)|\bm{\mathcal{F}}_{i-1}\right]\big\|\|\bar{{\bm{w}}}_{i-1}\|^2\nn
&\hspace{-0.1cm}\stackrel{(\ref{app35})}=\big\|I_M-2\bm{ \bar{D}}_{11,i}+\mathbb{E}\left[\bm D_{11,i}\bm D^{\sf{T}}_{11,i}|\bm{\mathcal{F}}_{i-1}\right]\big\|\|\bar{{\bm{w}}}_{i-1}\|^2\nn
&\hspace{-0.1cm}\stackrel{(\ref{app38})}=\big\|I_M-2\bm{ \bar{D}}_{11,i}+\bm {\bar{D}}^2_{11,i}+\bm R_{D_{11},i}\big\|\|\bar{{\bm{w}}}_{i-1}\|^2\nn
&\hspace{0.1cm}\leq\left(\big\|\left(I_M-\bm{ \bar{D}}_{11,i}\right)^2\big\|+\big\|\bm R_{D_{11},i}\big\|\right)\|\bar{{\bm{w}}}_{i-1}\|^2\nn
&\hspace{0.05cm}\stackrel{(c)}=\left(\big\|I_M-\bm{ \bar{D}}_{11,i}\big\|^2+\big\|\bm R_{D_{11},i}\big\|\right)\|\bar{{\bm{w}}}_{i-1}\|^2\nn
&\hspace{0.1cm}\leq\left((1-\sigma_{11}\mu_{\max})^2+\beta^2_{11}\mu_{\max}^2\right)\|\bar{{\bm{w}}}_{i-1}\|^2\label{app21}
\end{align}
where in step (a) we called upon the Rayleigh-Ritz characterization of eigenvalues \cite{Golub96,Johnson03}, and (b), (c) hold because $\|A\|=\lambda_{\max}(A)$ for any symmetric positive-semidefinite matrix $A$, and $\|A^2\|=\|A\|^2$ for any symmetric matrix $A$.

Computing the expectations again on both sides of (\ref{app21}), we have
\begin{align}\label{app44}
&\hspace{-0.3cm}\frac{1}{1-t}\mathbb{E}\left\{\mathbb{E}[\|(I_{M}-\bm D^{\sf{T}}_{11,i})\bar{{\bm{w}}}_{i-1}\|^2|\bm{\mathcal{F}}_{i-1}]\right\}\nn
&\leq\frac{1}{1-t}\left((1-\sigma_{11}\mu_{\max})^2+\beta^2_{11}\mu_{\max}^2\right)\mathbb{E}\|\bar{{\bm{w}}}_{i-1}\|^2\nn
&\stackrel{(a)}=\left(1-\sigma_{11}\mu_{\max}+\frac{\beta^2_{11}\mu_{\max}^2}{1-\sigma_{11}\mu_{\max}}\right)\mathbb{E}\|\bar{{\bm{w}}}_{i-1}\|^2\nn
&\stackrel{(b)}\leq\left( 1-\sigma'_{11}\mu_{\max}\right)\mathbb{E}\|\bar{{\bm{w}}}_{i-1}\|^2
\end{align}
where in step (a) we set $t= \sigma_{11}\mu_{\max}$, and in (b) positive number $\sigma'_{11}<\sigma_{11}$, and $\mu_{\max}$ is small enough such that
$\sigma'_{11}\leq\sigma_{11}-\left(1-\sigma_{11}\mu_{\max}\right)^{-1}{\beta^2_{11}}\mu_{\max}$.
We can now establish (\ref{equ28a}) by substituting (\ref{app44}) into (\ref{app021}), and completing the argument starting from Eq. (9.69) in the proof of Theorem 9.1 in \cite[pp. 516--521]{ASayed2014}, where the quantity $b={0}_{MN}$ (appeared in (9.54) of \cite{ASayed2014}).

We next establish (\ref{equ28b}). Compared to the proof for Theorem 9.2 in \cite{ASayed2014}, the main difference, apart from the second-order moments evaluated in (\ref{app44}), is the term \begin{equation}\label{app60}
\frac{1}{(1-t)^3}\mathbb{E}\left[\|(I_{M}-\bm D^{\sf{T}}_{11,i})\bar{{\bm{w}}}_{i-1}\|^4\right]
\end{equation}
for any $0<t<1$, which appeared in (9.117) of \cite{ASayed2014}.
Let
\begin{align}
\bm K_i &\triangleq \left(I_{M}-\bm D_{11,i}\right)\left(I_{M}-\bm D^{\sf{T}}_{11,i}\right)\bar{{\bm{w}}}_{i-1}\left(\bar{{\bm{w}}}_{i-1}\right)^{\sf{T}}\nn
&\hspace{0.5cm}\times\left(I_{M}-\bm D_{11,i}\right)\left(I_{M}-\bm D^{\sf{T}}_{11,i}\right)\label{app24}\\
\bm L_i &\triangleq \left(\left(I_{M}-\bm D_{11,i}\right)\left(I_{M}-\bm D^{\sf{T}}_{11,i}\right)\right)^2\label{app25}.
\end{align}
Then, both matrices $\bm K_i$ and $\bm L_i$ are symmetric positive semi-definite.
Thus, we have
\begin{align}
&\hspace{-2cm}\mathbb{E}\left[\|(I_{M}-\bm D^{\sf{T}}_{11,i})\bar{{\bm{w}}}_{i-1}\|^4|\bm{\mathcal{F}}_{i-1}\right]\nonumber\\
&\hspace{-1.2cm}=\left(\bar{{\bm{w}}}_{i-1}\right)^{\sf{T}}\mathbb{E}\left[\bm K_i|\bm{\mathcal{F}}_{i-1}\right]\bar{{\bm{w}}}_{i-1}\nonumber\\
&\hspace{-1.2cm}\leq\lambda_{\max}\left(\mathbb{E}\left[\bm K_i|\bm{\mathcal{F}}_{i-1}\right]\right)\left\|\bar{{\bm{w}}}_{i-1}\right\|^2\nonumber\\
&\hspace{-1.2cm}\stackrel{(a)}\leq\mathrm{Tr}\left(\mathbb{E}\left[\bm K_i|\bm{\mathcal{F}}_{i-1}\right]\right)\left\|\bar{{\bm{w}}}_{i-1}\right\|^2\nonumber\\
&\hspace{-1.2cm}=\left(\bar{{\bm{w}}}_{i-1}^{\sf T}\mathbb{E}\left[\bm L_i|\bm{\mathcal{F}}_{i-1}\right]\bar{{\bm{w}}}_{i-1}\right)\|\bar{{\bm{w}}}_{i-1}\|^2\nn
&\hspace{-1.2cm}\leq\lambda_{\max}\left(\mathbb{E}\left[\bm L_i|\bm{\mathcal{F}}_{i-1}\right]\right)\|\bar{{\bm{w}}}_{i-1}\|^4\nn
&\hspace{-1.2cm}=\|\mathbb{E}\left[\bm L_i|\bm{\mathcal{F}}_{i-1}\right]\|\|\bar{{\bm{w}}}_{i-1}\|^4\label{app26}
\end{align}
where the inequality (a) holds because $\lambda_{\max}(\Sigma)\leq \mathrm{Tr}(\Sigma)$ for any symmetric positive semi-definite matrix $\Sigma$. We proceed to deal with the term $\mathbb{E}\left[\bm L_i|\bm{\mathcal{F}}_{i-1}\right]$. Note that
\begin{align}\label{app27}
\bm L_i=I_M - \bm L_{1,i} + \bm L_{2,i} - \bm L_{3,i} +\bm L_{4,i}
\end{align}
where
\begin{align}
&\bm L_{1,i}&\hspace{-0.4cm}\triangleq&\hspace{0.1cm}2\bm D_{11,i}+2\bm D^{\sf{T}}_{11,i}\label{app51a}\\
&\bm L_{2,i}&\hspace{-0.4cm}\triangleq& \hspace{0.1cm}3\bm D_{11,i}\bm D^{\sf{T}}_{11,i}+\bm D^{\sf{T}}_{11,i}\bm D_{11,i}+\left(\bm D_{11,i}\right)^2+\left(\bm D^{\sf{T}}_{11,i}\right)^2\label{app51b}\\
&\bm L_{3,i}&\hspace{-0.4cm}\triangleq&\hspace{0.05cm}\left(\bm D_{11,i}\right)^2\bm D^{\sf{T}}_{11,i}+\bm D_{11,i}\left(\bm D^{\sf{T}}_{11,i}\right)^2+\nn
&&&\bm D_{11,i}\bm D^{\sf{T}}_{11,i}\bm D_{11,i}+\bm D^{\sf{T}}_{11,i}\bm D_{11,i}\bm D^{\sf{T}}_{11,i}\label{app51c}\\
&\bm L_{4,i}&\hspace{-0.4cm}\triangleq&\hspace{0.05cm}\left(\bm D_{11,i}\bm D^{\sf{T}}_{11,i}\right)^2\label{app51d}
\end{align}
and we have
$\mathbb{E}\left[\bm L_{1,i}|\bm {\mathcal{F}}_{i-1}\right]=4\bm{\bar{D}}_{11,i}$
according to (\ref{app35}).

Let $X$ be a constant matrix of size $M\times M$. Then,
\be\label{app48}
\mathbb{E}\left[\bm \Gamma_{k,i}X\bm \Gamma_{j,i}\right]=\left\{\begin{array}{cl}(1-r_k)(1-r_j)X,&k\neq j\\X',&k=j\end{array}\right.
\ee
where $X'$ has the same form as (\ref{equ46}), and we can further rewrite $X'$ as (\ref{equ51}).
It follows that:
\begin{align}\label{app52}
&\mathbb{E}\left[\bm D^{\sf{T}}_{11,i}\bm D_{11,i}|\bm {\mathcal{F}}_{i-1}\right]-\left(\bm{\bar{D}}_{11,i}\right)^2 \nn
&\hspace{0.1cm}=\sum_{k=1}^N\sum_{j=1}^Nq_kq_j\mathbb{E}\left[\bm \Gamma_{k,i}\bm {H}_{k,i-1}\bm {H}_{j,i-1}\bm \Gamma_{j,i}|\bm {\mathcal{F}}_{i-1}\right]-\nn
&\hspace{0.5cm}\sum_{k=1}^N\sum_{j=1}^Nq_kq_j(1-r_k)(1-r_j)\bm {H}_{k,i-1}\bm {H}_{j,i-1}\nn
&\hspace{0cm}\stackrel{(\ref{app48})}=\sum_{k=1}^N\sum_{j\neq k=1}^Nq_kq_j(1-r_k)(1-r_j)\bm {H}_{k,i-1}\bm {H}_{j,i-1}+\nn
&\hspace{0.3cm}\sum_{k=1}^Nq_k^2(1-r_k)^2\bm {H}_{k,i-1}^2+\sum_{k=1}^Nq_k^2(1-r_k)r_k\diag\{\bm {H}_{k,i-1}^2\}-\nn
&\hspace{0.3cm}\sum_{k=1}^N\sum_{j=1}^Nq_kq_j(1-r_k)(1-r_j)\bm {H}_{k,i-1}\bm {H}_{j,i-1}\nn
&\hspace{0.1cm}=\sum_{k=1}^Nq_k^2(1-r_k)r_k\diag\{\bm {H}_{k,i-1}^2\}.
\end{align}
Recall from (\ref{app53}) that $\{\bm {H}_{k,i-1}>0\}$. Then, $\{\bm {H}_{k,i-1}^2>0\}$ and $\{\diag\{\bm {H}_{k,i-1}^2\}>0\}$. Computing Euclidean norms on both sides of (\ref{app52}), we have
\begin{align}
&\big\|\mathbb{E}\left[\bm D^{\sf{T}}_{11,i}\bm D_{11,i}|\bm {\mathcal{F}}_{i-1}\right]-\left(\bm{\bar{D}}_{11,i}\right)^2\big\|\nn
&\hspace{0.3cm}\stackrel{(a)}\leq\sum_{k=1}^Nq_k^2(1-r_k)r_k\big\|\diag\{\bm {H}_{k,i-1}^2\}\big\|\nn
&\hspace{0.3cm}\stackrel{(b)}\leq\sum_{k=1}^Nq_k^2(1-r_k)r_k\left(\mathrm{Tr}\!\left[\bm {H}_{k,i-1}^2\right]\right)\nonumber\\
&\hspace{0.35cm}\leq\sum_{k=1}^Nq_k^2(1-r_k)r_k\left(M\lambda_{\max}\left(\bm {H}_{k,i-1}^2\right)\right)\nn
&\hspace{0.2cm}\stackrel{(\ref{app53})}\leq\sum_{k=1}^Nq_k^2(1-r_k)r_k\left(M\delta_d^2\right)\label{app54}
\end{align}
where in step (a) we used the property that $\|A+B\|\leq\|A\|+\|B\|$ \cite{Johnson03}, and (b) holds because $\|X\|\leq\mathrm{Tr}(X)$, for any symmetric positive semi-definite matrix $X$, and that $\mathrm{Tr}[\diag\{X\}]=\mathrm{Tr}[X]$.
Recall from (\ref{equ11}) that \cite[p. 509]{ASayed2014}
\be\label{app90}
q_k=\mu_k(e_k^{\sf T}A_2p)
\triangleq\mu_{\max}\tau_k(e_k^{\sf T}A_2p)
\ee
where $e_k$ denotes the $k$-th basis vector, which has a unit entry at
the $k$-th location and zeros elsewhere, and the parameter $\tau_k$ satisfies $\mu_k=\mu_{\max}\tau_k$.
Then, we have
\be\label{app91}
\big\|\mathbb{E}\left[\bm D^{\sf{T}}_{11,i}\bm D_{11,i}|\bm {\mathcal{F}}_{i-1}\right]-\left(\bm{\bar{D}}_{11,i}\right)^2\big\|=O(\mu_{\max}^2)
\ee
Likewise, it follows that
\begin{align}
\Big\|\mathbb{E}\left[\left(\bm D_{11,i}\right)^2|\bm{\mathcal{F}}_{i-1}\right]-\left(\bm{\bar{D}}_{11,i}\right)^2\Big\|&=O(\mu_{\max}^2)\label{app55a}\\
\Big\|\mathbb{E}\left[\left(\bm D_{11,i}^{\sf T}\right)^2|\bm{\mathcal{F}}_{i-1}\right]-\left(\bm{\bar{D}}_{11,i}\right)^2\Big\|&=O(\mu_{\max}^2).\label{app55b}
\end{align}
Recall from (\ref{app39}) and (\ref{app38}) that
\be\label{app56}
\big\|\mathbb{E}\left[\bm D_{11,i}\bm D_{11,i}^{\sf T}|\bm{\mathcal{F}}_{i-1}\right]-\left(\bm{\bar{D}}_{11,i}\right)^2\big\|=O(\mu_{\max}^2).
\ee
Substituting (\ref{app91})--(\ref{app56}) into (\ref{app51b}), we obtain
\be\label{app57}
\big\|\mathbb{E}\left[\bm L_{2,i}|\bm{\mathcal{F}}_{i-1}\right]-6\left(\bm{\bar{D}}_{11,i}\right)^2\big\|=O(\mu_{\max}^2).
\ee
Similarly, it can be verified that
\begin{eqnarray}
\big\|\mathbb{E}\left[\bm L_{3,i}|\bm{\mathcal{F}}_{i-1}\right]-4\left(\bm{\bar{D}}_{11,i}\right)^3\big\|=O(\mu_{\max}^3)\label{app58a}\\
\big\|\mathbb{E}\left[\bm L_{4,i}|\bm{\mathcal{F}}_{i-1}\right]-\left(\bm{\bar{D}}_{11,i}\right)^4\big\|=O(\mu_{\max}^4).\label{app58b}
\end{eqnarray}
It follows that
\begin{align}
\|\mathbb{E}\left[\bm L_{i}|\bm{\mathcal{F}}_{i-1}\right]\|&=
\|I-\mathbb{E}\left[\bm L_{1,i}|\bm{\mathcal{F}}_{i-1}\right]+\mathbb{E}\left[\bm L_{2,i}|\bm{\mathcal{F}}_{i-1}\right]-\nn
&\hspace{0.5cm}\mathbb{E}\left[\bm L_{3,i}|\bm{\mathcal{F}}_{i-1}\right]+\mathbb{E}\left[\bm L_{4,i}|\bm{\mathcal{F}}_{i-1}\right]\|\nn
&=\big\|I-4\bm{\bar{D}}_{11,i}+6\left(\bm{\bar{D}}_{11,i}\right)^2-4\left(\bm{\bar{D}}_{11,i}\right)^3+\nn
&\hspace{0.25cm}\left(\bm{\bar{D}}_{11,i}\right)^4+\left(\mathbb{E}\left[\bm L_{2,i}|\bm{\mathcal{F}}_{i-1}\right]-6\left(\bm{\bar{D}}_{11,i}\right)^2\right)\nn
&\hspace{0.25cm}-\left(\mathbb{E}\left[\bm L_{3,i}|\bm{\mathcal{F}}_{i-1}\right]-4\left(\bm{\bar{D}}_{11,i}\right)^3\right)\nn
&\hspace{0.25cm}+\left(\mathbb{E}\left[\bm L_{4,i}|\bm{\mathcal{F}}_{i-1}\right]-\left(\bm{\bar{D}}_{11,i}\right)^4\right)\big\|\nn
&\leq\big\|I-4\bm{\bar{D}}_{11,i}+6\left(\bm{\bar{D}}_{11,i}\right)^2-4\left(\bm{\bar{D}}_{11,i}\right)^3+\nn
&\hspace{0.25cm}\left(\bm{\bar{D}}_{11,i}\right)^4\big\|+\big\|\mathbb{E}\left[\bm L_{2,i}|\bm{\mathcal{F}}_{i-1}\right]-6\left(\bm{\bar{D}}_{11,i}\right)^2\big\|\nn
&\hspace{0.25cm}+\big\|\mathbb{E}\left[\bm L_{3,i}|\bm{\mathcal{F}}_{i-1}\right]-4\left(\bm{\bar{D}}_{11,i}\right)^3\big\|\nn
&\hspace{0.25cm}+\big\|\mathbb{E}\left[\bm L_{4,i}|\bm{\mathcal{F}}_{i-1}\right]-\left(\bm{\bar{D}}_{11,i}\right)^4\big\|\nonumber\\
&=\|\left(I-\bm{\bar{D}}_{11,i}\right)^4\|+O(\mu_{\max}^2)\nn
&=\|I-\bm{\bar{D}}_{11,i}\|^4+O(\mu_{\max}^2)\nn
&\hspace{-0.15cm}\stackrel{(\ref{app39})}\leq\left(1-\sigma_{11}\mu_{\max}\right)^4+O(\mu_{\max}^2).\label{app59}
\end{align}
Substituting into (\ref{app26}), and taking expectations again on both sides, we have
\begin{align}\label{app62}
&\frac{1}{(1-t)^3}\mathbb{E}\left[\|(I_{M}-\bm D^{\sf{T}}_{11,i})\bar{{\bm{w}}}_{i-1}\|^4\right]\nn
&\hspace{0.5cm}\leq\frac{1}{(1-t)^3}\left(\left(1-\sigma_{11}\mu_{\max}\right)^4+O(\mu_{\max}^2)\right)\mathbb{E}\|\bar{{\bm{w}}}_{i-1}\|^4\nn
&\hspace{0.45cm}\stackrel{(a)}=\left(1-\sigma_{11}\mu_{\max}+O(\mu_{\max}^2)\right)\mathbb{E}\|\bar{{\bm{w}}}_{i-1}\|^4\nn
&\hspace{0.5cm}\leq\left(1-\sigma''_{11}\mu_{\max}\right)\mathbb{E}\|\bar{{\bm{w}}}_{i-1}\|^4
\end{align}
for some positive constant $\sigma''_{11}<\sigma_{11}$, and for small enough $\mu_{\max}$, where in step (a) we set $t\triangleq\sigma_{11}\mu_{\max}$.
Then, the result in (\ref{equ28b}) can be obtained by continuing from Eq. (9.117) (by choosing $t = \sigma_{11}\mu_{\max}$) in the proof of Theorem 9.2 in \cite[pp. 523--530]{ASayed2014}.
\section{Proof of Theorem \ref{theo2}}\label{APP3}
Define
\begin{equation}\label{app21a}
\mathcal{F}\triangleq\mathbb{E}\left[\bm{\mathcal{B}}'_{i}\otimes_b\bm{\mathcal{B}}'_{i}\right]^{\sf T}
\end{equation}
Then, by following similar techniques shown in the proof of Lemma 9.5 \cite[pp. 542--546]{ASayed2014}, we have
\begin{equation}\label{equ37}
(I-\mathcal{F})^{-1}=[(p\otimes p)(\mathds{1}\otimes\mathds{1})^{\sf{T}}]\otimes Z^{-1}+O(1)
\end{equation}
where
\begin{equation}\label{equ38}
Z\triangleq\sum\limits_{k=1}^Nq_k(1-r_k)\left[\left(H_k\otimes I_{M}\right)+\left(I_{M}\otimes H_k\right)\right].
\end{equation}
The desired results (\ref{equ43}) and (\ref{equ72}) in Theorem \ref{theo2} now follow by referring to the proofs of Theorem 11.2 and Lemma 11.3 in \cite[pp. 583--596]{ASayed2014}, and Theorem 11.4 in \cite[pp. 608--609]{ASayed2014}.

Evaluating the squared Euclidean norms on both sides of (\ref{equ30}) and taking expectations conditioned on $\bm{\mathcal{F}}_{i-1}$, then
taking expectations again we get
\begin{multline}\label{app88}
\mathbb{E}[\|\widetilde{\bm{w}}_i'\|^2_{\mbox{\rm bvec}(I_{NM})}]= \mathbb{E}\left\{\|\widetilde{\bm{w}}_{i-1}'\|^2_{\mathcal{F}\mbox{\rm bvec}(I_{NM})}\right\}+\\\mathbb{E}\left\{\|\bm{s}_i\|^2_{\mathbb{E}\left[\left(\bm{\Gamma}_i\mathcal{M}\mathcal{A}_2\right)\otimes_b\left(\bm{\Gamma}_i\mathcal{M}\mathcal{A}_2\right)\right]\mbox{\rm bvec}(I_{NM})}\right\}
\end{multline}
where we used the weighted vector notation $\|x\|^2_{\sigma}=\|x\|^2_{\Sigma}$ with $\sigma=\mbox{\rm bvec}(\Sigma)$ and $\mbox{\rm bvec}(\cdot)$ denoting the block vector operation \cite[p. 588]{ASayed2014}.
Iterating the relation we get
\begin{multline}\label{app89}
\mathbb{E}[\|\widetilde{\bm{w}}_i'\|^2_{\mbox{\rm bvec}(I_{NM})}]= \mathbb{E}\left\{\|\widetilde{\bm{w}}_{-1}'\|^2_{\mathcal{F}^{i+1}\mbox{\rm bvec}(I_{NM})}\right\}+\\\sum_{n=0}^i\mathbb{E}\left\{\|\bm{s}_i\|^2_{\mathbb{E}\left[\left(\bm{\Gamma}_i\mathcal{M}\mathcal{A}_2\right)\otimes_b\left(\bm{\Gamma}_i\mathcal{M}\mathcal{A}_2\right)\right]\mathcal{F}^{n}\mbox{\rm bvec}(I_{NM})}\right\}
\end{multline}
where the first-term corresponds to a transient component that dies out with time, and the convergence rate of $\mathbb{E}\|\bm{\widetilde{w}}_{k,i}\|^2$
towards the steady-state regime is seen to be dictated by $\rho\left(\mathcal{F}\right)$ \cite[p. 592]{ASayed2014}. Now, let
\be
\Gamma\triangleq\mathbb{E}\bm \Gamma_i=\diag\{(1-r_k)\}_{k=1}^N\otimes I_M\label{app67a}
\ee
\be
\mathcal{M}'\triangleq\mathcal{M}{\Gamma}=\diag\{\mu_k(1-r_k)\}_{k=1}^N\otimes I_M\label{app67b}
\ee
\be
{\mathcal{B}'}\triangleq\mathbb{E}\bm{\mathcal{B}}'_{i}=\mathcal{A}_2^{\sf{T}}\left(I-\mathcal{M}'\mathcal{H}\right)\mathcal{A}_1^{\sf{T}}\label{app67c}
\ee
We now rewrite (\ref{app21a}) in terms of ${\mathcal{B}}'$ as
\begin{align}\label{app74}
\mathcal{F}\hspace{-0.1cm}\stackrel{(\ref{equ31})}=&\mathbb{E}\left[\left(\mathcal{A}_2^{\sf{T}}\left(I-\mathcal{M}\bm \Gamma_i\mathcal{H}\right)\mathcal{A}_1^{\sf{T}}\right)\otimes_b\left(\mathcal{A}_2^{\sf{T}}\left(I-\mathcal{M}\bm \Gamma_i\mathcal{H}\right)\mathcal{A}_1^{\sf{T}}\right)\right]^{\sf{T}}\nn
&\hspace{-0.48cm}=\left(\mathcal{A}_1\otimes_b\mathcal{A}_1\right)\big(I-I\otimes_b\left(\mathcal{H}\Gamma\mathcal{M}\right)-\left(\mathcal{H}\Gamma\mathcal{M}\right)\otimes_b I+\nn
&\mathbb{E}\left[\left(\mathcal{H}\bm \Gamma_i\mathcal{M}\right)\otimes_b\left(\mathcal{H}\bm \Gamma_i\mathcal{M}\right)\right]\big)\left(\mathcal{A}_2\otimes_b\mathcal{A}_2\right)
\nn
&\hspace{-0.48cm}=\left[{\mathcal{B}}'\otimes_b{\mathcal{B}}'\right]^{\sf T} + \Delta_F(\mu_{\max}^{2})
\end{align}
where $\Delta_F(\mu_{\max}^{2})$ is a matrix whose entries are in the order of $O(\mu_{\max}^{2})$.
Following similar techniques to the proof of Theorem 9.3 \cite[pp. 535--540]{ASayed2014}, we make the same Jordan canonical decomposition for matrix $\mathcal{P}=\mathcal{A}_1\mathcal{A}_2$ as (\ref{app6}), then substituting into (\ref{app67c}) we get
\be\label{app82}
{\mathcal{B}}'
=\left(\mathcal{V}_\epsilon^{-1}\right)^{{\sf{T}}}(\mathcal{J}^{{\sf{T}}}-{\mathcal{D}}'^{{\sf{T}}})\mathcal{V}_\epsilon^{\sf{T}}
\ee
where
\begin{align}\label{app83}
{\mathcal{D}}'^{\sf{T}}&\triangleq\mathcal{V}_\epsilon^{{\sf{T}}}\mathcal{A}_2^{{\sf{T}}}\mathcal{M}'\mathcal{H}\mathcal{A}_1^{{\sf{T}}}\left(\mathcal{V}_\epsilon^{-1}\right)^{\sf{T}}\nonumber\\
&=\left[\begin{array}{cc}{D}_{11}'^{\sf{T}}&{D}_{21}'^{\sf{T}}\\{D}_{12}'^{\sf{T}}&{D}_{22}'^{\sf{T}}\end{array}\right]
\end{align}
and
\be
{D}_{11}'=\sum\limits_{k=1}^Nq_k(1-r_k){H}_{k},\,
{D}_{21}'=O(\mu_{\max})\label{app84b}.
\ee
We now introduce the eigen-decomposition ${D}_{11}'^{\sf T} \triangleq U\Lambda U^{\sf T}$ for the symmetric positive-definite matrix ${D}_{11}'^{\sf T}$,
where $U$ is unitary, and $\Lambda$ is a diagonal matrix composed of the eigenvalues of ${D}_{11}'^{\sf T}$.
Let
\be\label{app79}
\mathcal{T} = \diag\{\mu_{\max}^{1/N}U,\mu_{\max}^{2/N}I_M,\ldots,\mu_{\max}^{(N-1)/N}I_M,\mu_{\max}I_M\}
\ee
then we have
\be\label{app71}
\mathcal{T}^{-1}\mathcal{V}_{\epsilon}^{\sf T}\mathcal{B}'\left(\mathcal{V}_{\epsilon}^{-1}\right)^{\sf T}\mathcal{T}=\left[\begin{array}{cc}B_{11}'&B_{12}'\\B_{21}'&B_{22}'\end{array}\right].
\ee
It follows that
$B_{11}'\triangleq I_M-\Lambda,\,B_{12}' = O(\mu_{\max}^{(N+1)/N})$ \cite[p. 538]{ASayed2014}.
The matrix $\mathcal{B}'$ has the same eigenvalues as the block matrix on the right hand side of (\ref{app71}).
By referring to Gershgorin's Theorem \cite{Golub96, Johnson03}, it is shown in \cite[pp. 539--540]{ASayed2014} that the union of the $M$ Gershgorin discs, each centered at an eigenvalue of $B_{11}'$ with radius $O(\mu_{\max}^{(N+1)/N})$, is disjoint from that of the other $M(N-1)$ Gershgorin discs, centered at the diagonal entries of $B_{22}'$, and therefore
\be\label{app73}
\rho\left(\mathcal{B}'\right)=\rho\left(B_{11}'\right)+O(\mu_{\max}^{(N+1)/N}).
\ee
Let
{\arraycolsep=1.4pt\def\arraystretch{1.5}
\begin{align}\label{app80}
\tilde{\Delta}_{F}&\triangleq\left(\mathcal{T}^{\sf T}\mathcal{V}_{\epsilon}^{-1}\right)\otimes_b\left(\mathcal{T}^{\sf T}\mathcal{V}_{\epsilon}^{-1}\right)\Delta_{F}(\mu_{\max}^2)\times\nn
&\hspace{0.4cm}\left(\mathcal{V}_{\epsilon}\left(\mathcal{T}^{-1}\right)^{\sf T}\right)\otimes_b\left(\mathcal{V}_{\epsilon}\left(\mathcal{T}^{-1}\right)^{\sf T}\right)\nn
&=\left[\begin{array}{c|ccc}O(\mu_{\max}^2)&&o(\mu_{\max}^{1/N})&\\
\hline
                                &O(\mu_{\max}^2)&&o(\mu_{\max}^{2/N})\\
                                o(\mu_{\max}^{2})&&\ddots&\\
                               &o(\mu_{\max}^{1+{1}/{N}})&&O(\mu_{\max}^2)\end{array}\right].
\end{align}}

\noindent It follows from (\ref{app80}) that all the diagonal blocks of $\tilde{\Delta}_{F}$ are in the order of $O(\mu_{\max}^2)$, the remaining block matrices in the first row are in the order of $o(\mu_{\max}^{1/N})$, the remaining block matrices in the first column are in the order of $o(\mu_{\max}^{2})$, and the upper and lower triangular blocks in the $(2,2)$th block of $\tilde{\Delta}_{F}$ are in the order of $o(\mu_{\max}^{2/N})$ and $o(\mu_{\max}^{1+1/N})$ respectively. Then, substituting (\ref{app71}) and (\ref{app80}) into (\ref{app74}), we have
\begin{align}\label{app75}
&\hspace{-0.3cm}\left(\mathcal{T}^{\sf T}\mathcal{V}_{\epsilon}^{-1}\right)\otimes_b\left(\mathcal{T}^{\sf T}\mathcal{V}_{\epsilon}^{-1}\right)\mathcal{F}\left(\mathcal{V}_{\epsilon}\left(\mathcal{T}^{-1}\right)^{\sf T}\right)\otimes_b\left(\mathcal{V}_{\epsilon}\left(\mathcal{T}^{-1}\right)^{\sf T}\right)\nn
&=\left(\left[\begin{array}{cc}B_{11}'&B_{12}'\\B_{21}'&B_{22}'\end{array}\right]\otimes_b\left[\begin{array}{cc}B_{11}'&B_{12}'\\B_{21}'&B_{22}'\end{array}\right]\right)^{\sf T}+\tilde{\Delta}_{F}\nn
&=\left[\begin{array}{cc}F_{11}&F_{12}\\F_{21}&F_{22}\end{array}\right]^{\sf T}
\end{align}
where
\be
F_{11}=B_{11}'\otimes B_{11}'+O(\mu_{\max}^{2}),\,
F_{12}=O(\mu_{\max}^{(N+1)/N})\label{app76b}.
\ee
Recall that
$B_{11}'$ is a diagonal matrix, so is $B_{11}'\otimes B_{11}'$, then we have
\be\label{app81}
\diag\{F_{11}\}=\diag\{\ \lambda\left(B_{11}'\otimes B_{11}'\right)\} + O(\mu_{\max}^{2})
\ee
which means that the diagonal entries of $F_{11}$ are the eigenvalues of $B_{11}'\otimes B_{11}'$ perturbed by a second-order term, $O(\mu^2_{\max})$.
Referring to Gershgorin's Theorem, the union of the $M^2$ Gershgorin discs, centered at the diagonal entries of $F_{11}$ with radius $O(\mu_{\max}^{(N+1)/N})$, is disjoint from the union of the other $M^2(N^2-1)$ Gershgorin discs, centered at the diagonal entries of $F_{22}$. Note that $\mathcal{F}$ has the same eigenvalues as the block matrix on the right hand side of (\ref{app75}), and that eigenvalues are invariant under a transposition operation.
It follows from (\ref{app73}) that
\be\label{app77}
\rho\left(\mathcal{F}\right)=\rho\left(B_{11}'\otimes B_{11}'\right)+O(\mu_{\max}^{(N+1)/N}).
\ee
Using the fact that
$\rho\left(B_{11}'\otimes B_{11}'\right)=\left[\rho\left(B_{11}'\right)\right]^2$,
we arrive at the desired result (\ref{equ88}).
\section{Examining the Difference in (\ref{equ64})}\label{APP6}
\begin{figure}[htbp]
\centering
\includegraphics[width=2.2in]{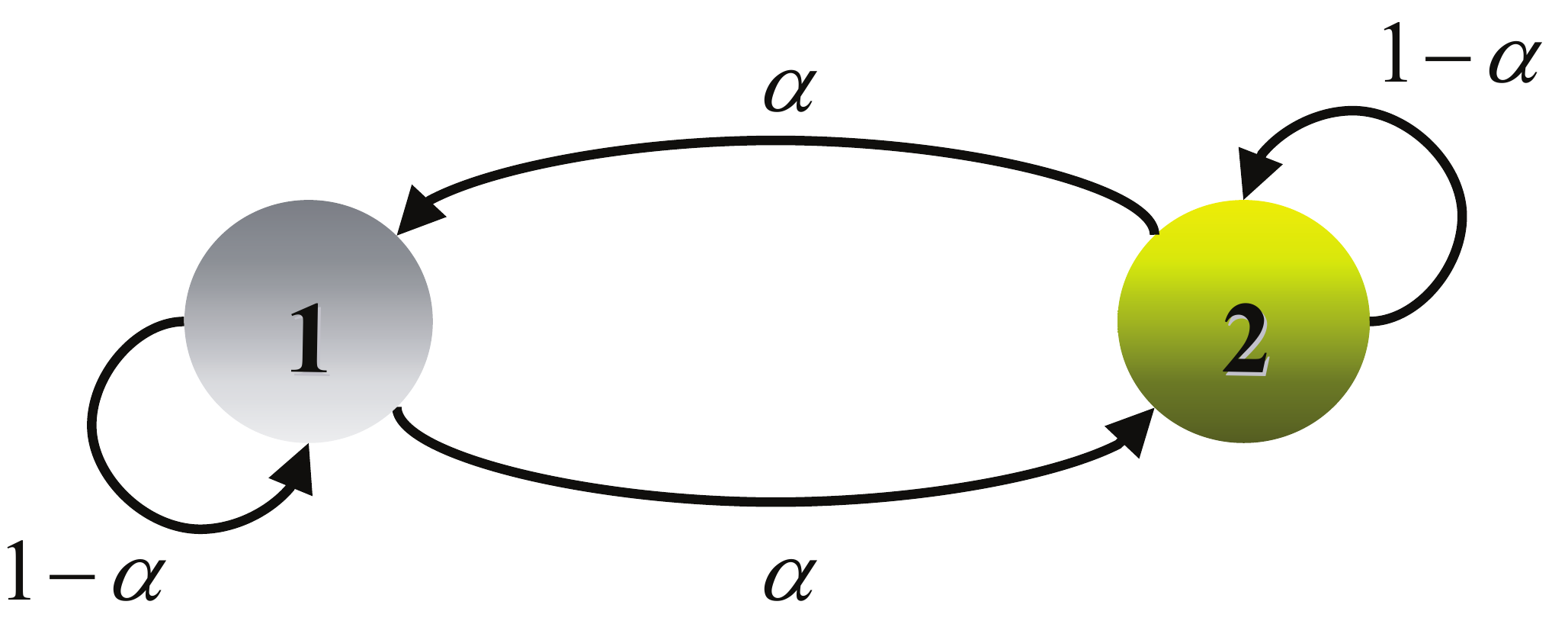}
\caption{A two-agent MSE network with a doubly-stochastic combination matrix.}
\label{two_agents_network}
\end{figure}

We revisit the MSE networks discussed in (\ref{equ48}).
Assume that there are only two agents in the network as shown in Fig. \ref{two_agents_network}, namely, $N = 2$, with $M = 2$.
Assume that
$\sigma_{v,1}^2>\sigma_{v,2}^2$.
For simplicity, uniform parameters $\{q_k\equiv q\}$ are used across the agents (which may occur, for example, in the ATC or CTA forms when the step-sizes are uniform across the agents, i.e., $\{\mu_k\equiv\mu\}$, and doubly-stochastic combination matrices are adopted. In this case, we get $\{q_k\equiv \mu/N\}$ \cite[pp. 493--494]{ASayed2014}). Let
\be\label{equ92}
R_{u,1}=\left[\begin{array}{cc}|\pi_1|&\pi_1\\ \pi_1&1\end{array}\right],\:R_{u,2}=\left[\begin{array}{cc}|\pi_2|&\pi_2\\ \pi_2&1\end{array}\right]
\ee
where the numbers $|\pi_1|<1,|\pi_2|<1$ ensure that $R_{u,1}>0,R_{u,2}>0$. Then, expression (\ref{equ64}) can be rewritten as
\begin{align}\label{equ83}
&\hspace{-0.2cm}\mathrm{MSD}_{\mathrm{coor},k}-\mathrm{MSD}_{\mathrm{grad},k}\nn
&\hspace{-0.2cm}\stackrel{(\ref{equ49b})}={r}q\mathrm{Tr}\Big(\left(R_{u,1}+R_{u,2}\right)^{-1}
\big(\sigma_{v,1}^2\left(\mbox{\rm diag}\{R_{u,1}\}-R_{u,1}\right)+\nn
&\hspace{0.5cm}\sigma_{v,2}^2\left(\mbox{\rm diag}\{R_{u,2}\}-R_{u,2}\right)\big)
\Big)\nn
&\hspace{-0.2cm}\stackrel{(\ref{equ92})}={r}q\mathrm{Tr}\Bigg(\left[\begin{array}{cc}|\pi_1|+|\pi_2|&\pi_1+\pi_2\\\pi_1+\pi_2&2\end{array}\right]^{-1}\times\nn
&\hspace{0.4cm}\left[\begin{array}{cc}0&-\pi_1\sigma_{v,1}^2-\pi_2\sigma_{v,2}^2\\-\pi_1\sigma_{v,1}^2-\pi_2\sigma_{v,2}^2&0\end{array}\right]\Bigg)\nn
&\hspace{0cm}=\frac{2rq(\pi_1+\pi_2)(\pi_1\sigma_{v,1}^2+\pi_2\sigma_{v,2}^2)}{2\left(|\pi_1|+|\pi_2|\right)-(\pi_1+\pi_2)^2}
\end{align}
where the denominator is positive
for all $|\pi_1|<1,|\pi_2|<1$. Then, $\mathrm{MSD}_{\mathrm{coor},k}<\mathrm{MSD}_{\mathrm{grad},k}$ if, and only if
$(\pi_1+\pi_2)(\pi_1\sigma_{v,1}^2+\pi_2\sigma_{v,2}^2)<0$,
which implies that
\begin{subequations}
\begin{empheq}[left=\empheqlbrace]{align}
0<\pi_2<1,-\pi_2<\pi_1<-\left(\sigma_{v,2}^2/\sigma_{v,1}^2\right)\pi_2\label{equ86}\\
-1<\pi_2<0,-\left(\sigma_{v,2}^2/\sigma_{v,1}^2\right)\pi_2<\pi_1<-\pi_2.\label{equ87}
\end{empheq}
\end{subequations}
Otherwise, $\mathrm{MSD}_{\mathrm{coor},k}\geq\mathrm{MSD}_{\mathrm{grad},k}$.
\section{Proof of Corollary \ref{cor1}}\label{APP4}
In the case when the matrices $\{H_k\}$ or $\{G_k\}$ are diagonal, it follows from (\ref{equ64}) and (\ref{equ89}) that $\mathrm{MSD}_{\mathrm{coor},k}-\mathrm{MSD}_{\mathrm{grad},k}=0$, which verifies (\ref{equ52h}).

More generally, according to (\ref{equ6}) and (\ref{app45e}), we have
\ben
\left(\sum\limits_{k=1}^Nq_kH_k\right)^{-1}>0,\:\,
\sum\limits_{k=1}^Nq_k^2G_k\geq0,\:\,
\sum\limits_{k=1}^Nq_k^2\mbox{\rm diag}\{G_k\}\geq0
\een
Then, applying the inequality \cite{Fang1994}:
\be\label{app40}
\lambda_{\min}(A)\mathrm{Tr}(B)\leq\mathrm{Tr}(AB)\leq\lambda_{\max}(A)\mathrm{Tr}(B)
\ee
for any symmetric positive semi-definite matrices $A$ and $B$, where $\lambda_{\min}(A)$ and $\lambda_{\max}(A)$ represent respectively the largest and smallest eigenvalues of $A$, we obtain
\begin{multline}\label{equ52d}
\mathrm{MSD}_{\mathrm{coor},k}-\mathrm{MSD}_{\mathrm{grad},k}
\leq\frac{r}{2}\Bigg\{\lambda_{\max}\left(\left(\sum\limits_{k=1}^Nq_kH_k\right)^{-1}\right)-\\
\lambda_{\min}\left(\left(\sum\limits_{k=1}^Nq_kH_k\right)^{-1}\right)\Bigg\}\sum\limits_{k=1}^Nq_k^2\mathrm{Tr}(G_k)
\end{multline}
where we substituted (\ref{equ89}) into (\ref{equ64}) and used the relation $\mathrm{Tr}(G_k)=\mathrm{Tr}\left(\diag\{G_k\}\right)$.
Then, noting that
\be\label{equ52b}
0<\sum\limits_{k=1}^N q_k\lambda_{\min}\left(H_k\right)
\leq\lambda\left(\sum\limits_{k=1}^Nq_kH_k\right)\leq\sum\limits_{k=1}^N q_k\lambda_{\max}\left(H_k\right)
\ee
we have
\begin{multline}\label{equ52c}
1\Big/\left(\delta_d\sum_{k=1}^Nq_k\right)\stackrel{(a)}\leq1\Big/\sum\limits_{k=1}^N q_k\lambda_{\max}\left(H_k\right)\leq\\
\lambda\left(\left(\sum\limits_{k=1}^Nq_kH_k\right)^{-1}\right)\leq\\
1\Big/\sum\limits_{k=1}^N q_k\lambda_{\min}\left(H_k\right)\stackrel{(b)}\leq1\Big/\left(\nu_d\sum_{k=1}^Nq_k\right)
\end{multline}
where the inequalities (a) and (b) hold due to (\ref{equ6}).
Substituting (\ref{equ52c}) into (\ref{equ52d}) gives the upper bound for the difference $\mathrm{MSD}_{\mathrm{coor},k}-\mathrm{MSD}_{\mathrm{grad},k}$ as shown by (\ref{equ52g}). Then, by following a similar argument, we obtain the lower bound for the difference as the opposite number of the upper bound, which leads to the desired result in Corollary \ref{cor1}.
\section{Proof of Corollary \ref{cor3}}\label{APP5}
We start from the MSD expression in (\ref{equ43a}) and note first that
\begin{align}\label{equ51c}
G_k'= (1-r_k)^2\left(G+\frac{r_k}{1-r_k}\mbox{\rm diag}\{G\}\right)
\end{align}
Substituting into (\ref{equ43a}) we have:
\begin{align}
\mathrm{MSD}_{\mathrm{coor},k}
=&\:\frac{1}{2}\left(\sum\limits_{k=1}^Nq_k(1-r_k)\right)^{-1}\left(\sum\limits_{k=1}^Nq_k^2(1-r_k)^2\right)\times\nn
&\:\mathrm{Tr}\left(H^{-1}\left(G+\frac{r_k}{1-r_k}\diag\{G\}\right)\right)\nn
=&\:\frac{1}{2}\left(\sum\limits_{k=1}^Nq_k(1-r_k)\right)^{-1}\left(\sum\limits_{k=1}^Nq_k^2(1-r_k)^2\right)\times\nn
&\:\mathrm{Tr}\left(H^{-1}G\right)+\frac{1}{2}\left(\sum\limits_{k=1}^Nq_k(1-r_k)\right)^{-1}\times\nn
&\:\left(\sum\limits_{k=1}^Nq_k^2(1-r_k)r_k\right)\mathrm{Tr}\left(H^{-1}\diag\{G\}\right)\nn
=&\:\frac{1}{2}\left(\sum\limits_{k=1}^Nq_k(1-r_k)\right)^{-1}\left(\sum\limits_{k=1}^Nq_k^2(1-r_k)^2\right)\times\nn
&\:\mathrm{Tr}\left(H^{-1}G\right)+
\frac{1}{2}(\theta-\alpha)\mathrm{Tr}\left(H^{-1}\diag\{G\}\right)\label{equ63a}
\end{align}
where (\ref{equ63a}) holds because
\begin{align}\label{equ70}
\theta-\alpha&=\frac{\sum_{k=1}^Nq_k^2(1-r_k)}{\sum_{k=1}^Nq_k(1-r_k)}-\frac{\sum_{k=1}^Nq_k^2(1-r_k)^2}{\sum_{k=1}^Nq_k(1-r_k)}\nn
&=\frac{\sum_{k=1}^Nq_k^2(1-r_k)r_k}{\sum_{k=1}^Nq_k(1-r_k)}
\end{align}
with the numbers $\alpha$ and $\theta$ being defined in (\ref{equ78}) and (\ref{equ79}), respectively. Recall that
\be\label{equ63b}
\mathrm{MSD}_{\mathrm{grad},k}=\frac{1}{2}\left(\sum\limits_{k=1}^Nq_k\right)^{-1}\sum\limits_{k=1}^Nq_k^2\mathrm{Tr}\left(H^{-1}G\right)
\ee
is the MSD performance for the full-gradient case. Thus,
\begin{align}\label{equ63c}
&\mathrm{MSD}_{\mathrm{coor},k}-\mathrm{MSD}_{\mathrm{grad},k}\nn
&\hspace{0.2cm}=\frac{1}{2}\left(\frac{\sum_{k=1}^Nq_k^2(1-r_k)^2}{\sum_{k=1}^Nq_k(1-r_k)}-\frac{\sum_{k=1}^Nq_k^2}{\sum_{k=1}^Nq_k}\right)\mathrm{Tr}\left(H^{-1}G\right)\nn
&\hspace{0.6cm}+\frac{1}{2}(\theta-\alpha)\mathrm{Tr}\left(H^{-1}\diag\{G\}\right)\nn
&\hspace{0.2cm}=\frac{\alpha }{2}\mathrm{Tr}\left(H^{-1}G\right)+\frac{1}{2}(\theta-\alpha)\mathrm{Tr}\left(H^{-1}\diag\{G\}\right).
\end{align}
Applying (\ref{app40}) and (\ref{equ6}) to (\ref{equ63c}), we obtain the desired results for the MSD performance shown in Corollary \ref{cor3}. The result for the ER performance in Corollary \ref{cor3} can be shown by subtracting the ER expression, $\mathrm{ER}_{\mathrm{grad},k}$, on the both sides of (\ref{equ74}).

\bibliographystyle{IEEEbib}
\bibliography{refs}

\begin{IEEEbiography}[{\includegraphics[width=1in,height=1.25in,clip,keepaspectratio]{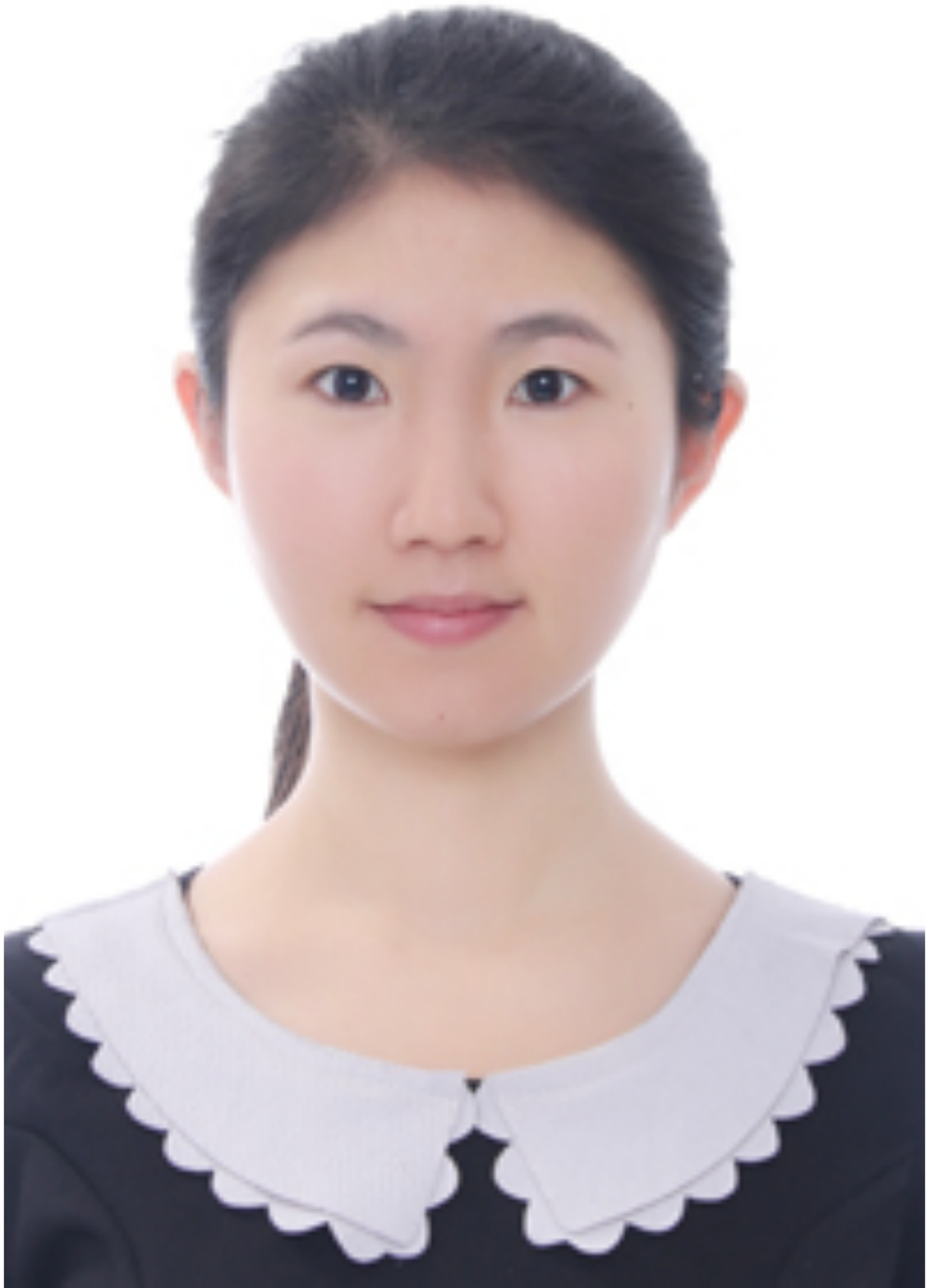}}]{Chengcheng Wang}
(S'17) received the B.Eng. degree in Electrical Engineering and Automation, and Ph.D degree in Control Science and Engineering, from the College of Automation, Harbin Engineering University, Harbin, China, in 2011 and 2017, respectively. She is currently a Research Fellow in the School of Electrical and Electronic Engineering, Nanyang Technological University, Singapore. From Sep. 2014 to Sep. 2016, she was a visiting graduate researcher in Adaptive Systems Laboratory at the University of California, Los Angeles, California, USA. Her research interests include adaptive and statistical signal processing, and distributed adaptation and learning.
\end{IEEEbiography}
\begin{IEEEbiography}[{\includegraphics[width=1in,height=1.25in,clip,keepaspectratio]{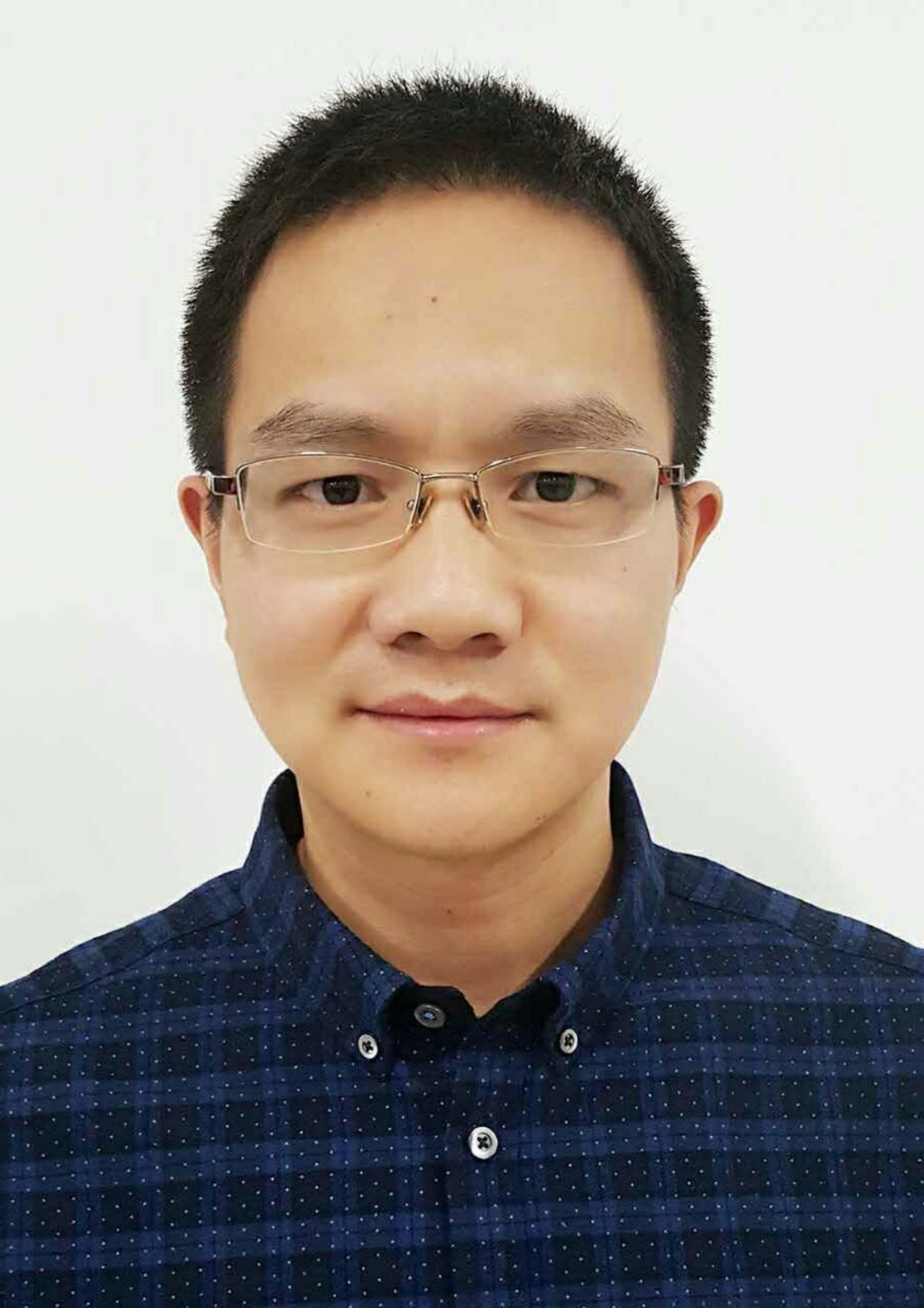}}]{Yonggang Zhang}
(S'06--M'07--SM'16) received the B.S. and M.S. degrees from the College of Automation, Harbin Engineering University, Harbin, China, in 2002 and 2004, respectively. He received his Ph.D. degree in Electronic Engineering from Cardiff University, UK in 2007 and worked as a Post-Doctoral Fellow at Loughborough University, UK from 2007 to 2008 in the area of adaptive signal processing. Currently, he is a Professor of Control Science and Engineering in Harbin Engineering University (HEU) in China. His current research interests include signal processing, information fusion and their applications in navigation technology, such as fiber optical gyroscope, inertial navigation and integrated navigation.
\end{IEEEbiography}
\vspace{-0.9cm}
\begin{IEEEbiography}[{\includegraphics[width=1in,height=1.25in,clip,keepaspectratio]{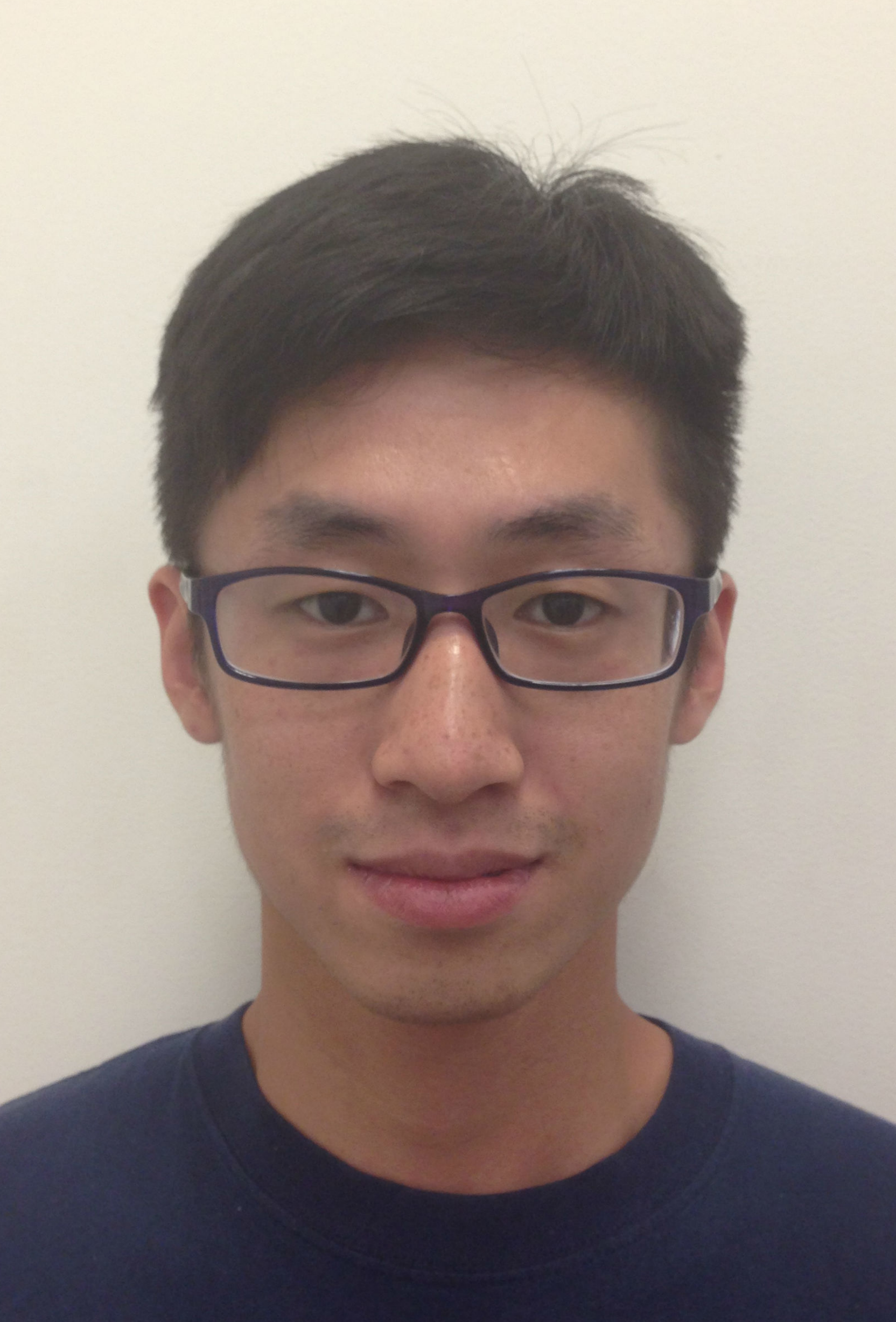}}]{Bicheng Ying}
(S'15) received his B.S. and M.S. degrees from Shanghai Jiao Tong University (SJTU) and University of California, Los Angeles (UCLA) in 2013 and 2014, respectively. He is currently working towards the PhD degree in Electrical Engineering at UCLA.
His research interests include multi-agent network processing, large-scale machine learning, distributed optimization, and statistical signal processing.
\end{IEEEbiography}
\vspace{-0.5cm}
\begin{IEEEbiography}[{\includegraphics[width=1in,height=1.25in,clip,keepaspectratio]{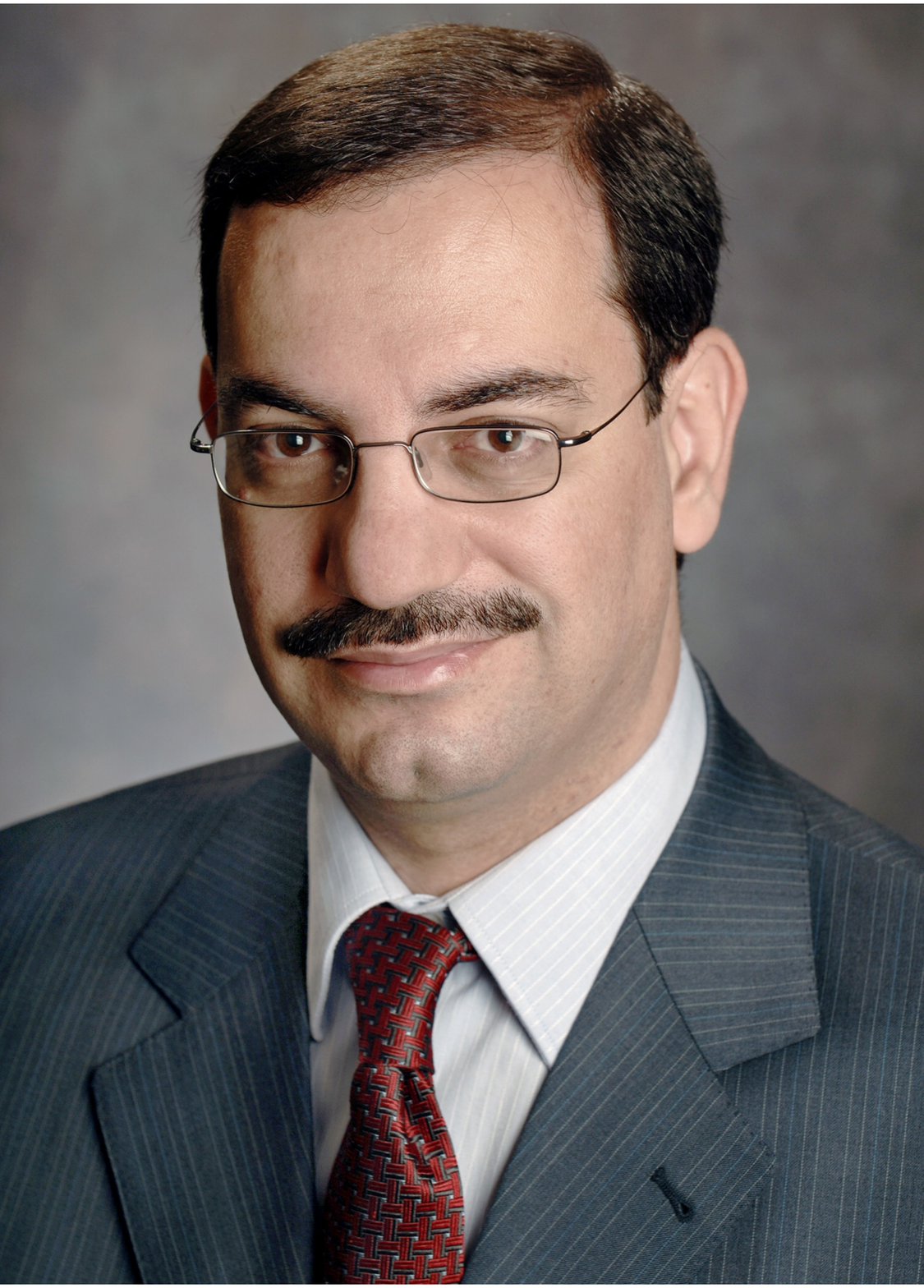}}]{Ali H. Sayed}
(S'90--M'92--SM'99--F'01) is Dean of Engineering at EPFL, Switzerland. He has also served as distinguished professor and former chairman of electrical engineering at UCLA. An author of over 500 scholarly publications and six books, his research involves several areas including adaptation and learning, statistical signal processing, distributed processing, network and data sciences, and biologically-inspired designs. Dr. Sayed has received several awards including the 2015 Education Award from the IEEE Signal Processing Society, the 2014 Athanasios Papoulis Award from the European Association for Signal Processing, the 2013 Meritorious Service Award, and the 2012 Technical Achievement Award from the IEEE Signal Processing Society. Also, the 2005 Terman Award from the American Society for Engineering Education, the 2003 Kuwait Prize, and the 1996 IEEE Donald G. Fink Prize. He served as Distinguished Lecturer for the IEEE Signal Processing Society in 2005 and as Editor-in-Chief of the IEEE TRANSACTIONS ON SIGNAL PROCESSING (2003--2005). His articles received several Best Paper Awards from the IEEE Signal Processing Society (2002, 2005, 2012, 2014). He is a Fellow of the American Association for the Advancement of Science (AAAS). He is recognized as a Highly Cited Researcher by Thomson Reuters. He is serving as President-Elect of the IEEE Signal Processing Society.
\end{IEEEbiography}
\end{document}